# Is methane the 'climate culprit'? The dangers of using imprecise, long-term GWP for methane to address the climate emergency.

Authors: Roger W. Bryenton[1]; Farrukh A. Chishtie[2]; T. Andrew Black[3], Mujtaba Hassan[4]; Tom Mommsen[5]; Devyani Singh[6].

## Abstract

The United Nations Environmental Program's (UNEP) Emissions Gap Report, 2023, "Temperatures hit new highs, yet world fails to cut emissions (again)", and in 2024, "No more hot air, emissions' massive gap between rhetoric and reality". A climate emergency has been declared yet policies and emission reductions continue to fail. Global temperature anomalies in recent years have not been modelled well.

Methane ($CH_4$) is a potent greenhouse gas (GHG) with a short atmospheric half-life (~8.4 years), and a perturbation lifetime of 11.8 ±1.8 yrs (IPCC AR6). It has a high, short-term impact on global warming: substantially greater than $CO_2$. Traditional metrics such as the 100-year Global Warming Potential ($GWP_{100}$) obscure the short-term, negative climatic effects of $CH_4$, potentially leading to inadequate policy responses. This study examines the limitations of $GWP_{100}$ in capturing the true, immediate climate impact of $CH_4$ and its inability to incorporate varying emissions, explores alternative metrics, and discusses the multi-facetted implications of this under-reporting of $CH_4$ emissions. Recalculation of 2024 Emissions Gap Report using a ten-year GWP of 105 increased $CH_4$'s warming effect to almost 90% of $CO_2$, rather than 25% using a $GWP_{100}$ of 28. We highlight the necessity of adopting a more immediate time horizon for $CH_4$'s warming effects, accelerating climate emergency action, while recognizing the adverse effects of the rapid growth rate of $CH_4$ emissions on reduction efforts.

To overcome the limitations of $GWP_{100}$, a static constant, we propose $GWP^{EFF}$ (*t*) which dynamically represents warming across various time periods. It is a novel, physically realistic measure that is simple to understand, and effective for policies in reducing short-term emissions such as $CH_4$. Policymakers can then adopt actions that pay attention to $CH_4$ as well as $CO_2$., and address the climate crisis with a better understanding of emission warming. The present, widespread use of incorrect steady-state assumptions in climate science (Cusworth, 2023; Mitloehner, 2020), where inputs equal outputs and are not highly variable, represents a systemic failure affecting both temperature predictions and policy effectiveness. Correcting this error enables accurate emissions tracking and modelling, and improved policy strategies to address the climate emergency.

<u>Summary Points for Policy Makers</u>:

- Increase emissions monitoring with timely, accurate, reporting to support stringent policies;
- Immediately develop policies to reduce methane emission from all sources with increased investment by government and private entities, and penalties for target failures;
- Urgently develop and implement policies committing to a planned phase-out of existing GHG emission sources, prioritized by the level of emissions, including existing and planned fossil fuel projects to be replaced by renewable energy sources and,
- Improve and expand communication among researchers, the public, and policy-makers.

---

[1] Independent researcher, PhD candidate, Vancouver, Canada; [2] PhD, Peaceful Society, Science, and Innovation Foundation, Vancouver, Canada; [3] PhD, Department of Soil Science, University of British Columbia, Canada; [4]PhD, Department of Space Science, Institute of Space Technology, Islamabad, Pakistan; [5]PhD, School of Environmental Studies, University of Victoria, Canada; [6]PhD, Independent Researcher, Vancouver, Canada. * Correspondence: Roger Bryenton (roger.bryenton@earthlink.net)

## Introduction

The 2023 and 2024 UNEP Emissions Gap Reports (EGR) and the latest 2024 World Meteorological Organization GHG Bulletin underscore the urgency of addressing climate change, noting that greenhouse gas (GHG) emissions reached new highs in 2022 and 2023, with global average temperatures rising as much as 1.8°C above pre-industrial levels (UNEP, 2023), and increasing further in 2024.

Despite years of research and temperature predictions, recent global temperatures have exceeded modelled values, prompting two years of EGR's expressing concern about increasing temperatures and GHG emissions. Resulting weather extremes have not only increased temperatures, but also caused serious health issues from heat domes; extreme wildfire activity in Canada, the US, Europe, Brazil and Australia; extreme rainfall events and flooding; more intense and greater numbers of tornados and hurricanes; and highly variable storms. Modelling and temperature predictions have not been accurate: the target of a 1.5 °C global increase was exceeded in 2024 (WMO, 2025) Mitigation efforts for emissions reductions were similarly unsuccessful as noted by the EGR's. It is clear that one or more factors have not been correctly included in predicting temperature and weather extremes, and that improved understanding of temperature anomalies is needed, particularly for emitted gases and their impact on warming. The Emissions Gap Reports from 2023 and 2024 identified the recent temperature extremes and the necessity of reducing emissions, as well as the science of emissions: comparing gases and emission warming factors. What was not covered was the high variability of methane emissions, nor the use of very short term, immediate $CO_2$ – equivalents. This is warming from short-lifetime gases such as methane which has a much greater warming effect than $CO_2$ despite low concentrations.

**ATMOSPHERIC THEORY, MODELLING, AND GAPS**

**Global Temperatures and Extremes** - The last four years were the warmest on record: 2022 was 0.86° C above the 20th century average of 13.9 °C; 2023 at 1.18 °C; and 2024 at 1.55 °C (WMO) exceeding the IPCC target of 1.5 °C. Copernicus (EU) reported 2024 as 1.6 °C: "Each of the past 10 years was one of the 10 warmest years on record", while a "new record high was reached for daily global average temperature… 22 July 2024 at 17.16 °C" (WMO), the basis for Hausfather's (Berkeley Earth) comment "a bit warmer… than anticipated".

**Atmospheric Methane -** $CH_4$ is highly reactive with the hydroxyl radical (OH), which shortens its lifetime compared to $CO_2$. Since pre-industrial times, the methane concentration has more than doubled owing to anthropogenic emissions in the atmosphere (Staniaszek, 2022). This trend has significantly increased in recent years (Nisbet et al. 2019), particularly after 2020 (Staniaszek, 2022). The abilities of simple climate models are limited in their ability to accurately simulate atmospheric methane processes, especially the effect on the oxidizing capacity in the troposphere (Hayman, 2021). However, recent studies (Allen, 2021; Folberth, 2022; Heimann, 2020) have shown that Global Climate Models (GCMs) and the Aerosol and Chemistry Model Intercomparison Project (AerChemMIP) of CMIP6 are more capable of representing methane involving atmospheric processes over longer time horizons. A global time-varying precalculated methane surface concentration was used as a lower boundary condition (LBC) in these complex models (Staniaszek, 2022). We observed that CMIP models use annual concentrations of GHG's, which are unable to accommodate shorter time horizons for calculations.

Shindell (2005) used an emissions-driven configuration to simulate changes in atmospheric composition in response to increased methane emissions from the preindustrial to the present day in the tropospheric coupled chemistry-aerosol GCM of the Goddard Institute for Space Studies (GISS). He et al. (2020) employed the Geophysical Fluid Dynamics Laboratory Atmospheric Model (GFDL AM4.1), a methane emission-driven version, to reproduce the historical period by improving the methane emission component calculation. Staniaszek (2022) used a methane emissions-driven version of the UK Earth System Model (UKESM1) and explored the role of anthropogenic methane in the Earth system under the SSP3-7.0 scenario, which is the most extreme future methane trajectory in the CMIP6 GCMs, again using longer term data and calculations.

Studies such as those by Allen (2021), Folberth (2022), and Staniaszek (2022) demonstrated that the interactions of methane with atmospheric chemistry, including its reactivity with the hydroxyl radical (OH), are critical for accurately simulating its climate impacts. These models use a methane emission-driven approach to capture the dynamic feedback mechanisms that influence the atmospheric concentration of methane. As highlighted in these models, the significant radiative efficiency of methane suggests that even small fluctuations in its concentration can lead to substantial changes in global temperature. Therefore, there is an urgent need for a methane emission-driven treatment to simulate changes in the feedback mechanism on full-scale Earth system impacts; investigations planned by this team.

**Global Greenhouse Gas Concentrations and Rates of Increase and Decay** - The principal global warming GHGs, $CO_2$ and $CH_4$ have atmospheric concentrations of approximately 426 ppm (parts per million) for $CO_2$ and 1923 ppb (parts per billion) or 1.92 ppm, one 200$^{th}$ as much for $CH_4$. $CO_2$ is increasing at 2.4 ppm or 0.5% per year. Methane concentration increased at a range of 9 to 17 ppb in 2021, almost 1% per year; double that of $CO_2$. The 2023 increase was lower at 8.51 ppb. $CO_2$ has a long lifetime, decaying very slowly, while methane decays approximately 8.4% per year. The high rate of decay was identified as a potential major factor in GHG accounting. This rate of change was not reported by other researchers nor was the increase in the rate of change, discussed below.

$CO_2$ Emissions - Annual global fossil $CO_2$ emissions, as per EGR, 2024, Table 2.1 were 39.1 Gt.
$CH_4$ Emissions – Annual global total methane emissions are from 575 Tg (Mt) (calculated bottom-up) to 669 Tg (Mt) calculated top-down (Global Methane Budget, 2025). Averaging these provides a mid-range value of 663 Tg (Mt). Anthropogenic methane emissions are from to 331 Tg(Mt) to 343 Tg(Mt), with a mid-range of 337 Tg (Mt), or 54% of total. This agrees closely with the Emissions Gap Report, Table 2.1 value of 350 Mt, or 9.8Gt $CO_2$-e using a conversion of 28 for a 100 year warming factor. When corrected for immediate, short-term GWP of 120, the effective warming increases to 42 Gt $CO_2$–e, exceeding $CO_2$.

**Emissions Variability** – With highly variable emissions, the use of long-term warming over an extended time horizon, such as 100 years, may not accurately assess the timing or magnitude of warming. This is particularly relevant for methane with high emissions variability, both hourly and seasonally, and with rapid decay involved, described below for Barrow, Alaska and Mauna Loa, Hawaii

**Methane Decay and Replacement** - The decay rate of methane is approximately 8.4% per year, with a half-life of 8.4 years and a perturbation life of 12 years (IPCC AR5 and AR6). Nisbet et al. (2023) calculated the ratio of $CH_4$ emission rate to concentration rise to be 2.77 Mt of methane per parts per billion (ppb). This has profound implications for global annual emissions reporting, as it is unclear if methane decay and replacement is reported in various emissions reporting publications, discussed in the following sections. Total anthropogenic and natural global annual methane emissions are up to 669 Mt in a top-down analysis, above, (Global Methane Budget, 2025) of which 561 Mt is decay and replacement methane, almost 84%.

**Warming and Time Horizon** - Emissions data and $CO_2$ "equivalency" values ($CO_2$-e) for $CO_2$ and $CH_4$ from a literature review provide different "warming factors" for methane, depending upon the time horizon chosen which we applied to published emission reports, as "corrections". Most time horizons for assessing global warming, as used by IPCC in determining and reporting emissions, are based on 100 years, such as used in Canada, while some researchers use up to 500 years. (Fuglestvedt, 2000). We determined that these time horizons are not useful for short or very short lifetime gases or to determine their impacts, such as immediate to 5 or 10 years.

**Global Warming Potential (GWP) and Global Temperature Potential (GTP) -** The GWP metric, developed under the Kyoto Protocol and adopted by the IPCC, is defined as the time-integrated radiative forcing of a pulse emission over a specified time horizon relative to $CO_2$. GWP is used to compare the impact of different GHGs on climate forcing by converting their emissions into $CO_2$ equivalents ($CO_2$-e). The most commonly used time horizon is 100 years ($GWP_{100}$), which assigns a value of 25-28 to methane; 25 is most commonly used, conforming to the IPCC's AR4 and AR5 conventions, while 28 is used for AR6 methodologies. (Balcombe, 2018; Howarth, 2014). GTP is the resulting global temperature at a future time. Short-term, over ten years, GWP and GTP are similar because the planet has not absorbed the energy increase, particularly by oceans, Figure 1.

The GWP metric has several limitations.

1**. Time Horizon Selection:** The choice of time horizon greatly affects the GWP value, Figure 1.

> Methane's GWP is much higher over shorter time horizons: 120 in the immediate, one year time (Balcombe, 2018, and Howarth, 2021); 105 over 10 years; and 86 over 20 years (Shindell, Howarth, and Hughes), reflecting its intense short-term impact, (Howarth, 2014).

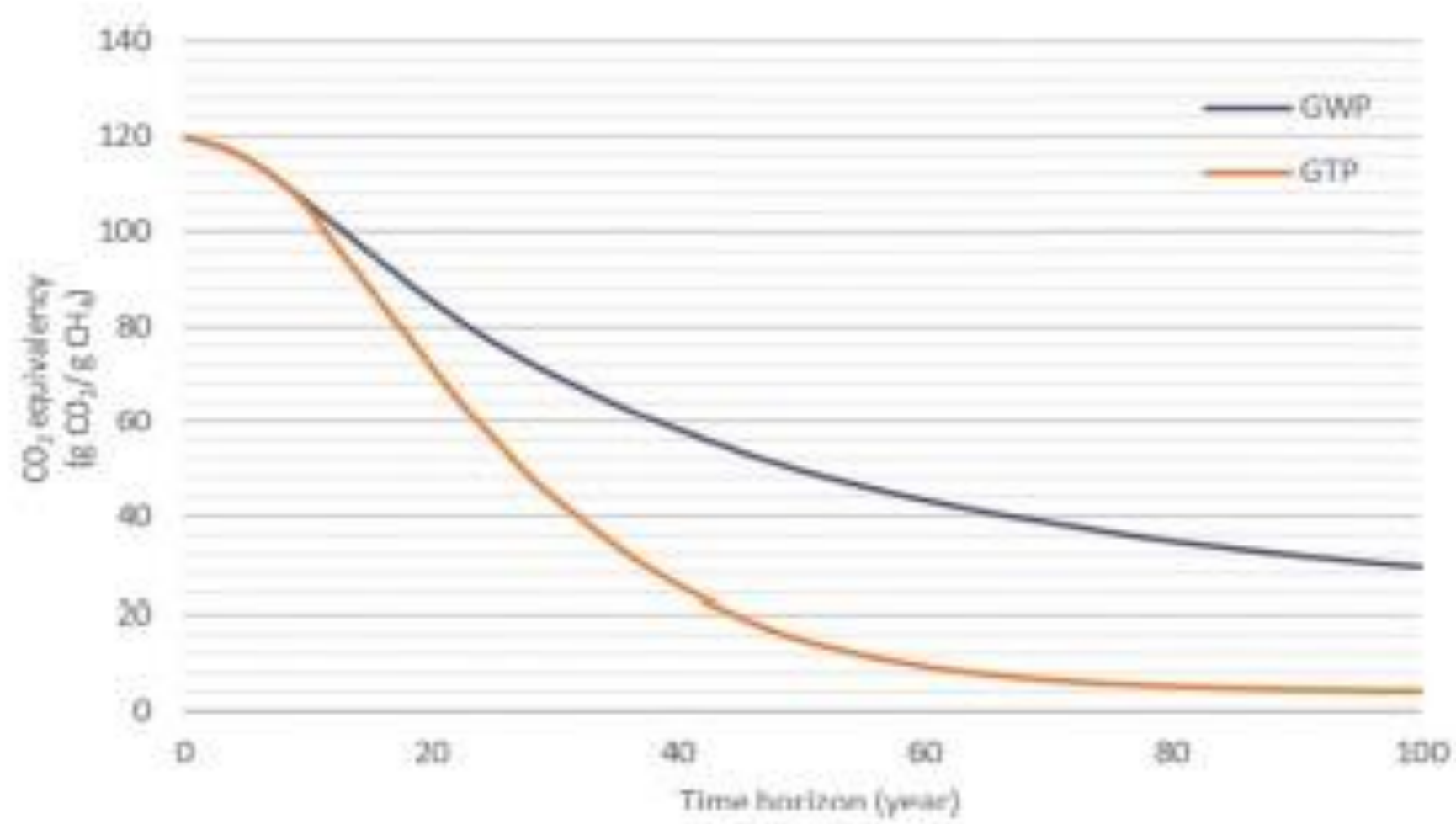

Figure 1. $CO_2$ equivalence of methane over different time horizons. GWP and GTP are equal for 10 years (Balcombe et al, 2018). GTP, Global Temperature Potential, is a measure of the temperature change caused by $CH_4$ at the end of the time period, relative to that for $CO_2$ which for very short terms, up to 10 years, is equivalent to the short term GWP, of 120 times $CO_2$.

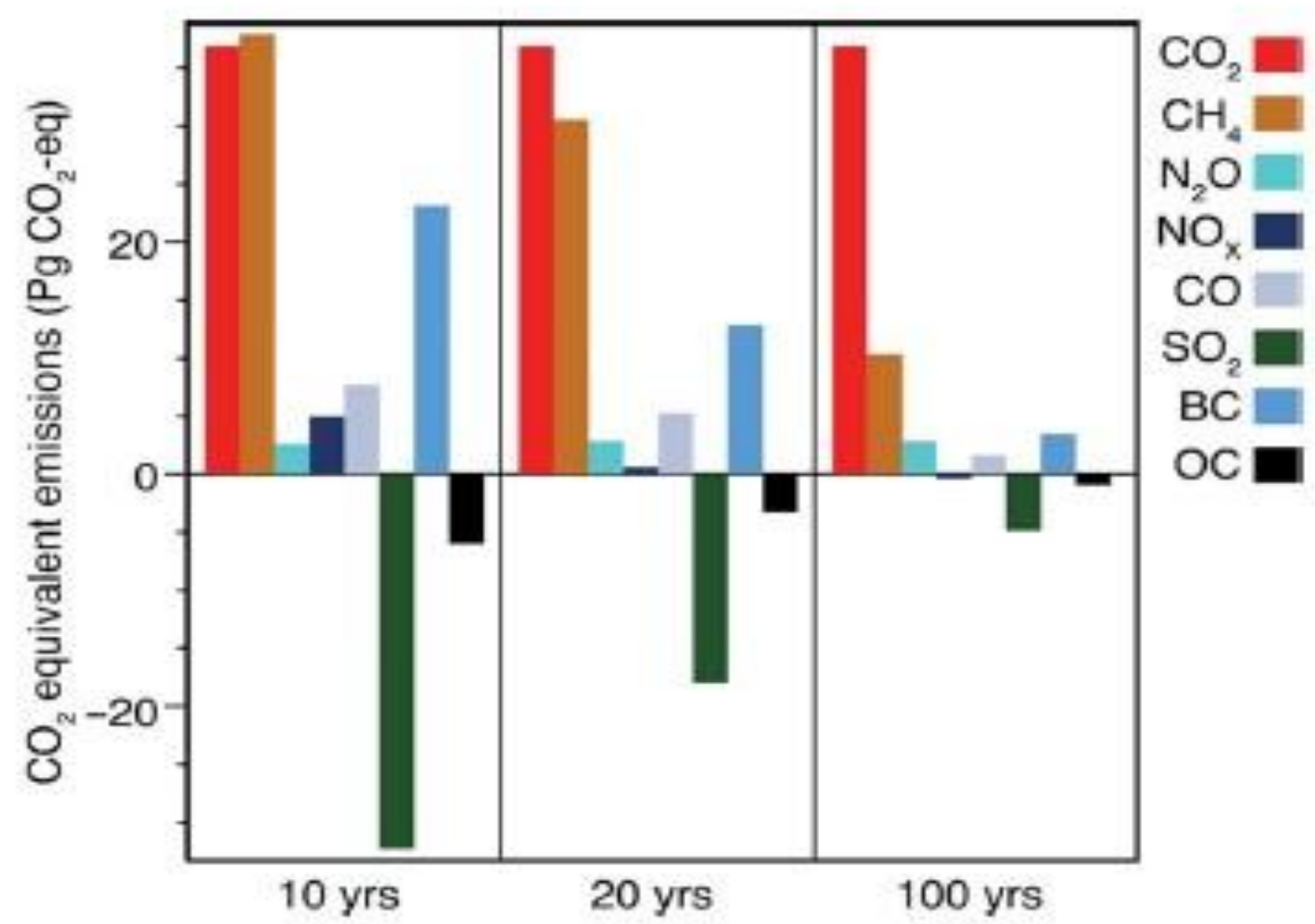

Figure 2. Over a ten-year horizon, (left panel), warming from methane emissions (second, brown bar) exceeds that from $CO_2$ emissions, (first, red bar) (Howarth et al, 2014).

**2. Physical Basis:** GWP measures the radiative forcing rather than direct temperature change, which can misrepresent the immediate climate impact of methane emissions, as shown in Figure 1. Note that the Global Temperature Potential (GTP) is equivalent to GWP for years one through ten (Balcombe, 2018).

**3. Pulse Emissions Focus:** GWP is based on single, pulse emissions, not accounting for sustained or variable emissions, especially non-linear and increasing emission profiles typical of many GHG sources (Balcombe, 2018), thus substantially under-report warming effects. Because GWP assumptions are based on linear, steady state flows, or a single pulse, they cannot be used as a metric for highly variable, temperature-related flows in the short-term.

**GWP* as an Alternative Metric** – In addition to $GWP_{100}$, recent studies have proposed alternative metrics such as GWP*, which critique that GWP is a model-based entity rather than a metric (Allen, 2018; Lynch, 2020). GWP* aims to capture the temperature implications of short-lived pollutants, such as methane, more accurately (Lynch, 2020; Meinshausen, Nicholls, 2022). GWP* scales GWP by its time horizon and considers emission rate changes over time, thus providing a more accurate representation of the impact on global warming, longer-term. It is a unified and refined representation of all GHG's, and while GWP* offers theoretical advantages, it also introduces significant variability and may be inconsistent with existing climate policy frameworks (Meinshausen, Nicholls, 2022). This study found that other researchers have used time horizons for GWP and GWP* varying from 50 to 500 years, which we contend cannot provide useful guidance for short-term or immediate warming factors, particularly for methane.

For longer term evaluations, 20 years or more, GWP* holds considerable opportunity for warming potential assessment, but is not suited to help determine the best immediate or "emergency response". This is particularly applicable to the choice of oil and methane gas extraction: to vent or to flare the gas. Venting causes an immediate warming, or GWP of 120, compared to flaring, which has a GWP of 1 due to the combustion of methane and $CO_2$ as the resultant emission, with lower warming.

**Research Objectives -** 1. Assess GWP metrics for the suitability of short-term temperature response to methane; 2. Examine methane's short-term variability; and 3. Develop a refined metric, $GWP^{EFF}$, suited to short-lived methane, for a more accurate $CO_2$ equivalence.

## Methods

**LITERATURE REVIEW**

Current climate research literature was assessed for applicability, particularly short-term emissions and warming. The IPCC convention of using a 100-year time horizon is not relevant in addressing immediate emissions reductions. Metrics such as the Global Warming Potential (GWP), GWP* and improvements were assessed for their applicability and limitations, particularly for very short-term time horizons less than 20 years, as were the necessary emissions reductions and policies to limit global warming.

**DATA EXTRACTION AND CRITICAL APPRAISAL**

Data were obtained from NOAA for temperatures and changes, and for $CO_2$ and $CH_4$ emissions at multiple locations, to assess variability on hourly, daily, seasonal and annual bases for applicability to modelling and warming metrics. The use of GWP and GTP were compared for very short-term time horizons of immediate to several years vs longer time horizons.

Examinations of emissions and temperature data for non-linearity and hourly, daily and annual dynamics were performed, using data from Mauna Loa, Hawaii, and from the Arctic/sub-Arctic area, Barrow, Alaska, with known extreme emissions. The acceleration of GHG emission rates were analyzed, as were the decay rates of $CO_2$ and $CH_4$ and the impact on increasing concentrations of GHG's.

**$GWP^{EFF}$ as a Simple Metric –** Given the complexities and under-reporting of short-term methane emissions by the use of GWP and GWP*, we developed a simplified approach, $GWP^{EFF}$, to address the deficiencies of GWP and GWP*.

**POLICY REVIEW AND IMPLICATIONS**

Greenhouse gas reduction policies were reviewed and implications assessed, given the failure to substantially reduce emissions in many countries, particularly Canada and British Columbia. Proposed revisions and improved policies are presented, as is the urgency of implementing effective policies.

# Results

**KEY FINDINGS**

1. Methane emissions are non-linear and highly dynamic, particularly in the Arctic. Hourly variations can exceed 11% of concentration. Over five years, Arctic concentration variations exceeded 20%. Rate of change of the increase of $CH_4$ was up to 15% per year. When combined with under-reporting of warming effects, the above factors result in substantial errors in determining warming effects over short time horizons.
2. Methane emissions and concentrations are accelerating, not just increasing. The Arctic rate of increase was 13% over five years.
3. Warming effects of methane have been substantially under-reported, by as much as 4.3 times.
4. Carbon dioxide emissions are moderately variable. Hourly are from 2.3% of concentration in MLO (Hawaii) to 8% in BRW (Alaska). Seasonally, $CO_2$ at MLO peaks April through June, with rapid declines to September. BRW peaks from September through December, with lows in July and August. Annually, MLO was 3% while BRW was 6%.
5. Methane emissions are much more variable. Hourly are from 5% of concentration (MLO) to 11.5% in BRW. Seasonally, $CH_4$ at MLO peaks fall-winter: BRW late summer. Annually, MLO was 4.8% and BRW was 8.4% in 2020, to 17% in 2024, a dramatic increase in four years.
6. Due to the long life of $CO_2$, annual (or monthly) values have provided reasonable estimates of global warming over longer time horizons.
7. Short-term determination of warming from methane cannot use GWP values based on 100-year time horizons.
8. $GWP^{EFF}$ based upon short-time horizons for methane provided more accurate warming factors.
9. Our detailed methane analysis demonstrates continuous accumulation for nearly five decades, not an immediate steady-state.

10. Policy Implications. With warming under-reported, the impact of methane has been substantially under-estimated, resulting in non-optimum policies and increasing global temperatures.

From our research, we contend that:

1. The short-term impact of methane is much greater than the generally accepted GWP of 100 years, which uses a warming factor of 25 to 27.9 (IPCC, AR5 and AR6) to report the climate impact of methane. This is significantly lower than the actual immediate warming factor of 120 (Balcombe, 2018; Howarth, 2021).
2. The conventional assumption is that methane emissions are linear with respect to time, either a pulse or flow, and do not increase in a non-linear or exponential manner (Cusworth, 2023).
3. Anthropogenic emissions are of concern, and while "natural" methane emissions are not under direct human control, they are often omitted from nationally reported emissions.
4. Little attention, if any, has been given to the non-linear effects that anthropogenic emissions may have on "natural" emissions which eventually result in irreversible warming feedback (IPCC AR6).

**GLOBAL WARMING IMPACTS AND POLICIES**

We show that the IPCC convention of using a 100-year time horizon is not relevant in addressing immediate emissions reductions. Importantly, the need for practical, non-academic approaches to conveying GHG reductions and the impacts on warming was recognized, based upon the failure to both accurately predict temperature increases, and to reduce emissions. The goal of our research is to both better understand GHG emissions, and to present them in a simple-to-understand form, enabling the development of better policy options and the optimal application of resources in GHG reductions. GWP and GTP, Figure 1, were compared for very short-term time horizons of immediate to several years vs longer time horizons. We determined that the high variability of methane emissions results in a mismatch between the use of longer-term modelling of air temperatures vs actual hourly, daily and monthly temperature changes.

**CLIMATE SCIENCE CONCEPTS, ANALYSES, AND LIMITATIONS**

A fundamental issue underlying temperature prediction failure is the widespread misrepresentation of methane's atmospheric behaviour in climate models, using concepts of GWP* with flow and stock emissions components (Allen, 2018). Many current approaches assume methane reaches immediate steady-state conditions (Mitloehner, 2020; Cusworth, 2023), neglecting the multi-decadal accumulation process and rising concentrations, confirmed by IPCC AR6 data. This modeling error, exemplified in recent publications (Fig. 3, Cusworth, 2023), systematically underestimates methane's near-term climate impact and misdirects policy prioritization from necessary urgent reductions (Ocko et al, 2021).

Our analysis demonstrates that correcting this fundamental misunderstanding of methane dynamics helps to explain recent temperature anomalies that have exceeded model predictions.

GWP and GWP* metrics have several limitations as they do not address the short-term, Figure 1. IPCC

emissions reporting protocol reinforces this limitation, by using a 100-year time horizon. GWP measures the radiative forcing rather than direct temperature change, which can misrepresent the immediate climate impact of methane emissions, shown in Figures 1 and 2. GWP is based on single, pulse emissions, not accounting for sustained or variable emissions. GWP* incorporates sustained flows, however it does not address non-linear, increasing emission profiles typical of many GHG sources (Balcombe, 2018), again substantially under-reports warming effects. Because GWP assumptions are based on linear, steady state flows, or a single pulse, they cannot be used as a metric for highly variable, temperature-related flows in the short term.

Thus, the emissions reporting for methane using 100 years for domestic waste, decomposition of forest harvesting waste, agriculture, or oil and gas extraction, was found to substantially under-estimate the warming impact over short time horizons (Howarth, 2014; Balcombe,2018). Immediate warming from methane can be as much as 120 times greater than $CO_2$ (Howarth, 2014; Balcombe, 2018), thus the 100-year value under-reports the actual warming effect by over four times.

The Emissions Gap Report (EGR, 2024 for anthropogenic $CH_4$ emissions of 9.8 Gt $CO_2$ –e used a GWP of 28. When adjusted for the short-term effect, this becomes 42 Gt, an increase of almost 4 times.

**Methane Decay Rate** - The high rate of decay of methane is a major factor in GHG accounting. Methane's atmospheric concentrations of approximately 1923 ppb (parts per billion) or 1.92 ppm, is one 200$^{th}$ that of $CO_2$. $CO_2$ is increasing at 2.4 ppm or 0.5% per year. Methane's concentration increased at a range of 9 to 17 ppb in 2021, almost 1% per year; double that of $CO_2$. Methane decays approximately 8.4% per year. With a concentration of 1923 ppb, the decay will be 165 ppb/year, more than eight times the annual increase which is less than 20 ppb; or less than 1%. Thus a minor increase in the decay rate will appear to substantially decrease the concentration, increasing warming, not reported by other researchers. Nor have they reported the increase in the rate of change, of 1% per year, nor the acceleration of the rate of increase, described below.

**POLICY IMPLICATIONS OF UNDER-REPORTED METHANE EMISSIONS**

We find that existing policies to reduce GHG emissions using under-reported emissions analyses have not and cannot provide optimum solutions. The largest sources of methane may not receive the largest reduction efforts: oil, gas and coal mining activities, anthropogenic waste, reducing wildfire-related methane emissions or agricultural sources from ruminants and rice production.

Policies and efforts to reduce emissions from $CO_2$ sources may overlook greater benefits of methane reduction efforts. Increased methane emissions from "natural" sources due to warming, can potentially be reduced indirectly, if anthropogenic sources are reduced, leading to reduced warming. Failure to capture the urgency of methane mitigation for near-term temperature stabilization can result. Climate, health, and financial benefits from methane reduction policies are not achieved.

**Broader Scientific Implications -** Policy ineffectiveness: Current strategies fail to address the most immediate warming drivers. Climate emergency acceleration: accumulated methane creates warming momentum not captured in steady-state models.

**SIMPLIFYING AND IMPROVING THE SHORT-TERM WARMING CONCEPT: $GWP^{EFF}$**

**"Effective Warming" -** We developed a simplified warming concept: $GWP^{EFF}$. It is based upon Howarth (2014) and Balcombe's (2018) work on short time horizons and $CO_2$ equivalency. It is simple and effective for policy purposes, as it avoids stocks and flows complexities of GWP and GWP*, and arbitrary coefficients. It is practical, easy to understand and directly applicable to methane reduction policies. It uses the "first year" GWP, a GWP (and GTP) value of 120 times the warming effect of $CO_2$, which an immediate release of methane creates. It represents the atmospheric response, such as occurs when methane is vented at gas production facilities, or is emitted from wildfires. It conveys the importance and immediacy of reduction to minimize warming effects and is useful for the public and for evaluation of policy options, given the need for urgent and immediate methane reductions, without requiring advanced emissions calculations.

**Mathematical Framework of $GWP^{EFF}$ –** We define the effective Global Warming Potential ($GWP^{EFF}$) to provide a time-sensitive measure of warming impact that incorporates atmospheric decay as indicated in Figure 1:

$$GWP^{EFF}(t) = GWP_i \times \exp\left(\frac{-(t-i)}{\tau}\right) \qquad \text{Eq (1)}$$

where $GWP_i$ = Warming potential at the reference time horizon i (1, 10, or 20 years), $t$ = Current time since emission, $i$ = Reference time horizon (1, 5, or 10 years), and $\tau$ = Atmospheric lifetime of the greenhouse gas. At the reference time horizon ($t = i$), $GWP^{EFF}(i) = GWP_i$, providing the benchmark warming potential for that specific timeframe. The units of $GWP^{EFF}$are called "warming units".

**Uncertainty Bounds** - The immediate GWP value of 120 carries significant uncertainty. The Greenhouse Gas Management Institute (2024) notes that "the uncertainty on these methane GWP values is roughly ±40%, or ±11 GWP points" for standard GWP values. IPCC assessments acknowledge that GWP "values should not be considered exact" and "when quoting a GWP it is important to give a reference to the calculation". For the immediate GWP value of 120, we estimate uncertainty of ±25%, based on variations in atmospheric chemistry models and radiative forcing calculations. This uncertainty range (90-150) is substantially lower than the factor-of-four systematic underestimation inherent in using $GWP_{100}$.

**Policy Application -** $GWP^{EFF}$ enables direct comparison of methane reduction strategies by converting all emissions to "warming units", using Eq (1), a simplified and practical metric where 1 warming unit represents the immediate temperature impact of 1 tonne $CO_2$. For example:

- the annual $CO_2$ increase of 2.4 ppm x 1 = 2.4 warming units.
- the annual $CH_4$ increase of .017 ppm x 120 = 2.04 warming units.

Thus $CH_4$ annual increase contributes 85% of $CO_2$'s immediate warming, despite being 140 times less concentrated.

**Comparative Analysis of Methane Atmospheric Models -** To validate our approach, we conducted a systematic comparison of methane atmospheric behaviour under different modeling assumptions:

1. Steady-State Assumption (used in conventional models): Methane concentration remains constant from year one, implying 100% annual replacement.
2. IPCC AR6 Dynamics (this study): Methane accumulates over 47 years following exponential approach to equilibrium, based upon improvements to Cusworth's paper (2023), Figure 3.
3. Policy-Relevant Timeframes: Impact assessment over 1–20-year horizons.

This comparative framework enables direct assessment of how different atmospheric assumptions affect climate impact calculations and policy recommendations.

**Correcting the Steady-State Fallacy: A One-Box Model Analysis of Methane Atmospheric Dynamics -** Figure 3 (Cusworth, 2023) contains a critical misrepresentation of methane atmospheric dynamics.

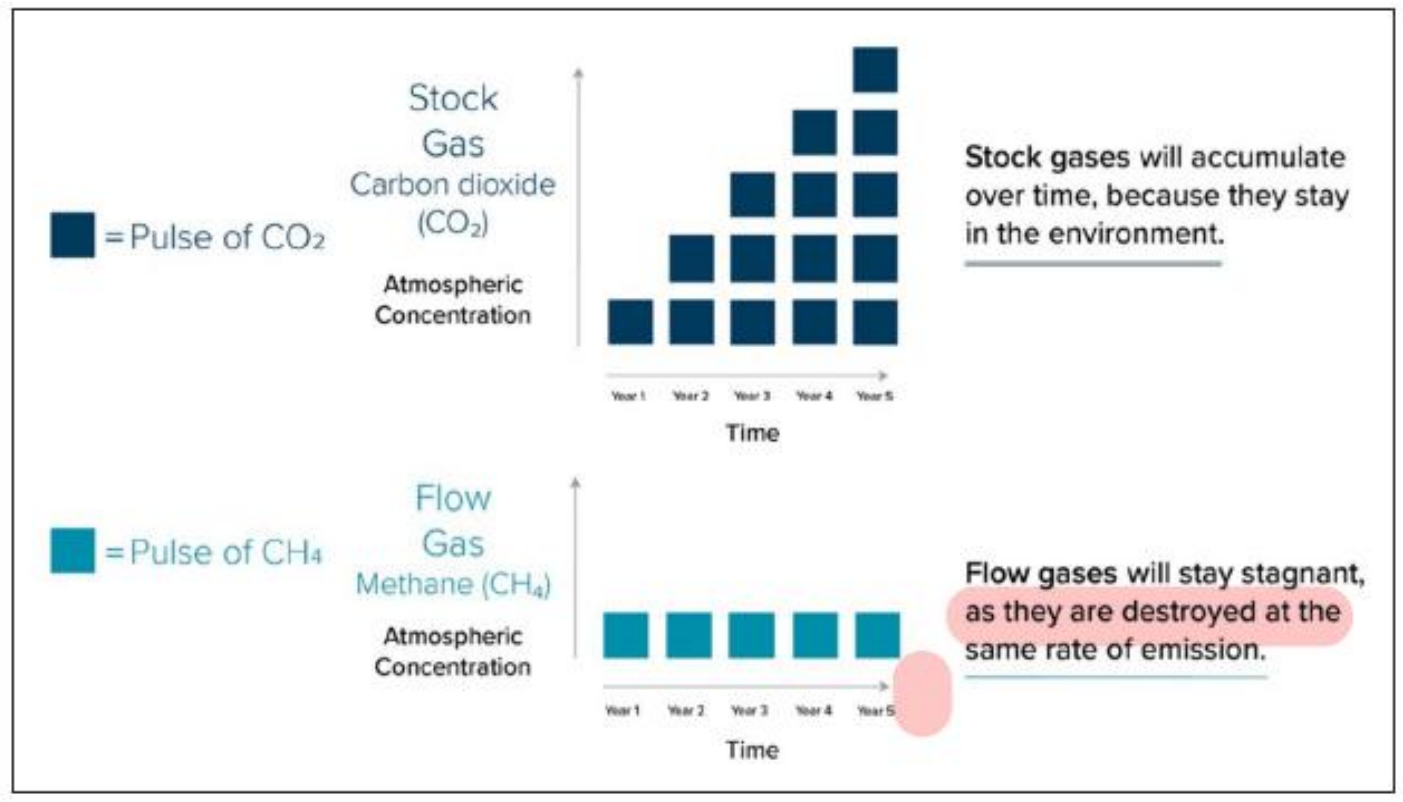

Figure 3. Cusworth's Figure 1: "stock" and "flow" of $CO_2$ and $CH_4$, with immediate equilibrium.

Figure 3 suggests that methane emissions reach immediate equilibrium, with the annotation stating, "gases will stay stagnant, as they are destroyed at the same rate of emission." This assumption is scientifically incorrect and contradicts established atmospheric chemistry principles.

However, following Jacobs (1999) and established atmospheric chemistry literature, steady-state conditions require multiple atmospheric lifetimes to achieve. We demonstrate this by showing methane's atmospheric behavior follows the one-box model, mass balance equation:

$$\frac{dm}{dt} = E - km \quad \text{Eq (2)}$$

where emissions *m* represents the total atmospheric mass of methane at time *t*, *E* are balanced against first-order loss *L* = *km*. For constant emissions, and an initial condition *m*(0) = 0, the solution of the differential equation is:

$$m(t) = \frac{E}{k}[1 - \exp(-kt)] \quad \text{Eq (3)}$$

Using IPCC AR6 data ($\tau$ = 11.8 years, k = 1/ $\tau$ = 0.0847 year$^{-1}$), our model approaches steady state exponentially over multiple decades. The following table demonstrates methane accumulation for constant unit emissions, showing continuous accumulation for nearly five decades.

| Year | Atmospheric Load | Annual Decay | Net Accumulation | % of Steady State |
|---|---|---|---|---|
| 1 | 0.96 | 0.08 | 0.92 | 8.1% |
| 5 | 4.08 | 0.35 | 0.65 | 34.5% |
| 10 | 6.74 | 0.57 | 0.43 | 57.1% |
| 20 | 9.64 | 0.82 | 0.18 | 81.6% |
| 30 | 10.88 | 0.92 | 0.08 | 92.1% |
| 47 | 11.59 | 0.98 | 0.02 | 98.1% |

Table 1. Computations from solution to Equation 3.

Methane requires approximately 47 years (~4 atmospheric lifetimes) to reach 98% of steady state, not immediate equilibrium. Current atmospheric conditions (1,923 ppb and rising at ~10 ppb/year) demonstrate ongoing disequilibrium. As Jacobs (1999) established, steady-state assumptions are only valid when conditions remain constant for periods much longer than the atmospheric lifetime, a condition clearly not met for methane. The steady-state fallacy leads to systematic underestimation of methane's warming impact and incorrect assessment of emission reduction benefits. The atmosphere is not a simple flow-through system for methane but a complex reservoir with substantial residence times that must be properly accounted for in climate impact assessment. This demonstrates continuous accumulation for nearly five decades, not an immediate steady-state as Cusworth contends.

**Quantitative Impact of the Error** – The climate impact is an underestimation when models assume an immediate steady-state (Cusworth, 2023), and the assumption of the methane burden at 1.0 units, results in a climate impact of 1 x $GWP_1$ = 120 warming units. Under actual methane accumulation (IPCC AR6 dynamics) by year 10 the burden is 6.896 units and the climate impact is 6.896 x $GWP_{1}$= 827 warming units. This results in an underestimation factor of 6.9 times.

**Broader Scientific Implications -** This error appears to be widespread in climate modeling literature, potentially explaining several factors:

1. Temperature prediction failures: In 2024, models underestimated near-term warming of 1.55 $^{\circ}$C by neglecting methane accumulation. The target temperature had been 1.5 $^{\circ}$C.
2. Policy ineffectiveness: Current strategies fail to address the most immediate warming drivers, such as wildfires and forest harvesting waste decomposition.
3. Climate emergency acceleration: Accumulated methane creates warming momentum not captured in steady-state models.

**Correction Requirements -** Climate models must incorporate:

1. Multi-decadal methane accumulation based on IPCC AR6 100-year lifetime data.
2. Dynamic emission scenarios rather than steady-state assumptions.
3. Policy-relevant timeframes (1-20 years) for methane impact assessment.
4. Feedback mechanisms between temperature increases and methane emissions.

The correction of the steady-state, fundamental error is essential for accurate climate prediction and effective policy development.

**Policy Implications of the Error -** The Cusworth steady-state assumption leads to serious under-estimation and policy failures as follows:

1. Systematic underestimation of methane's climate impact by factors of 2-10 depending on timeframe.
2. Misallocation of climate resources toward long-term $CO_2$ strategies while neglecting immediate methane reductions.
3. Failure to capture the urgency of methane mitigation for near-term temperature stabilization.
4. Incorrect assessment of the climate benefits from methane reduction policies.

**REVISIONS TO EMISSIONS DATA, DECAY, VARIABILITY ANALYSES AND POLICY IMPLICATIONS**

**Re-evaluating Methane's Impact** – We determined that using a 100-year warming factor is misleading and under-reports immediate and short-term warming effects of methane. The 2024 EGR report for 2023 emissions was 9.8 Gt of methane at a $CO_2$ e of 28. Using an immediate $CO_2$-e factor of 120, this becomes 42 Gt of $CO_2$ –e, exceeding the $CO_2$ warming of 39Gt. Thus the 100-year warming factor substantially under-estimates the short-term impacts of methane by a factor of 4.3, which can lead to policy measures that fail to address urgent climatic changes effectively (Ocko et al, 2021).

This recalculation also increases the 2023 global total emissions "equivalent" from 57.1 Gt $CO_2$ – e to 87 Gt $CO_2$-e, a 55 % rise, highlighting methane's dominant role as a warming agent, surpassing $CO_2$ .

**Decay rate of $CH_4$ -** This study identified the decay rate, and necessary $CH_4$ replacement to maintain increasing concentrations, as an important consideration not identified by other researchers, or incorrectly assessed (Cusworth, 2023), previously described. We believe that emissions variability and high decay rate helps explain recent, variable changes in apparent methane growth rates, annual concentrations, and accelerating global warming. With the rapid decay of methane, the replacement flow of 160 ppb is an order of magnitude greater than the annual increase. Thus, small changes in the decay rate can flip the apparent growth rate from an increase to a decrease.

**Global Atmospheric Methane Quantity** - The Global Methane Budget, 2020, shows methane emissions were estimated from 685 Tg(Mt)/year (top down) to 608 Mt/year (bottom up), a 13% difference. Averaging these results is 646.5 Tg (Mt)/year total. Anthropogenic sources are 382 Tg, or 60%. Decay is from 602 (bottom up), to 538 (top down) Tg (Mt) of $CH_4$, for an average of 570 Mt/year. As $CO_2$-e, that becomes 570 x 28 = 15.96 Gt of CO2-e. With global $CO_2$ emissions of 34.8 Gt, $CH_4$ is 46%. However, at a $GWP_1$ of 120, the corrected number is 570 x 120 = 68.4 Gt $CO_2$–e, double that of $CO_2$.

The annual methane increase, was from 36 to 21 Tg, (a 50% variation) with an average of 28.5 Tg (Mt). At 2.77 Tg/ppb, this implies an annual increase of 7.89 ppb, approximately agreeing with observed values. With total annual methane emissions of 646.5 Tg and decay of 604 Tg, that difference is 42.5 Tg or 11.7 ppb which agrees with the annual increase of 12.4 ppb.

**Global Methane Decay and Methane Budget Impact -** Using a half-life of 12 years for methane, the decay rate is 8.4% per year. The total anthropogenic and natural global annual methane decay emissions are thus 160 ppb/year, which, using 2.77 Tg (Mt) per ppb (Nisbet, 2023) for an annual methane decay replacement of 2.77 X 160 = 443 Mt. This does not agree with the Global Methane Budget, of 570 Mt as a sink, a difference of 127 Mt. It is close to the low end of the range of 496 Mt (Bottom Up) and 503 Mt (Top Down), but indicates a difference of 22%. We suggest that additional sinks may be under-reported, such as forest sequestration, and that the decay rate may be lower than estimated, given known variability of emissions.

Data analysis revealed a substantial acceleration of methane emissions and concentration globally. Methane global concentration, between 2000 and 2024 increased at a range of from 9 to 22 ppb/yr, almost 1% per year, with higher rates being in recent years. Some years showed annual declines in concentrations: 2000, 2004 and 2006.

**Global Methane Short-Term Variability** - We identified highly variable global methane concentration changes: hourly, daily, and monthly; with increases and decreases much greater than annual changes. In 2024 the annual increase was 9.13 ppb, while the September monthly increase was 9.62 ppb. The three-month increase was 21.04 ppb, more than double the annual increase. Monthly decreases were not as dramatic, in the order of 1/3 to ½ annual values. Seasonally, September-October was the highest concentration, while June and July were low as were December and January (Supplement).

These global methane findings support our premise that long-term horizons and warming factors do not facilitate useful data for short-term warming effects from methane, particularly for policy purposes. This is further supported by our findings from specific sites.

**Arctic and Mid-Latitude Emissions, Concentrations and Variability Analyses** - To further support our analyses in determining substantial global emissions variability, $CO_2$ and $CH_4$ emission data were accessed for two locations, one in the Arctic at Barrow (BRW), Alaska, the other at Mauna Loa (MLO), on Maui, Hawaii. This was to assess variability on hourly, daily, seasonally and annual bases including an examination of non-linearity: also hourly, daily and annually. The two sites incorporated high latitude and lower latitude regions, both in areas with low populations and anthropogenic emission sources. These were selected as potential "indicative" sites for potentially variable GHG emissions. The variability of $CO_2$ and $CH_4$ at Mauna Loa, Hawaii, and Barrow Alaska, was analyzed for time periods of 2000 to 2024, with a subset for detail and recent changes, 2020 to 2024 as Figure 4 below shows (See Supplemental Section).

**Data Locations -** At mid-latitude, Mauna Loa (MLO), Hawaii, is located at an upper elevation, 3397 m, representing well-mixed gases, away from population centres. The second, Barrow, (BRW) Alaska provides data from the Arctic/Subarctic zone, at a low elevation of 8 m, away from population, and in a known zone of rapidly increasing emissions and summer temperatures.

**Concentrations and Variability** –
**$CO_2$**. MLO, 2000 to 2016, annual average $CO_2$ concentration increased at an average of 2.3 ppm/yr or 0.5% per year. 2016 to 2025 increase was 2.5 ppm/yr with little variation from year to year. BRW, 2010 to 2025, annual average $CO_2$ concentration increased at a similar rate, 2.36 ppm/yr.

**Relevance -** GWP and GWP* are able to accurately represent emissions warming over time frames of 20 years or greater, due to assumed relatively stable emissions rates. However, determining a short-term "warming" potential using GWP or GWP* may involve considerable temporal errors. Hourly calculations of warming would be appropriate for modelling temperatures; however, this would far exceed the capabilities of most policy makers.

The variability of $CO_2$ is less important for warming compared to methane, due to methane's much greater warming effect over short-term time horizons: daily to seasonally. Figure 4, MLO, $CO_2$ emissions have considerable variability, with highs in late spring, May, declining quickly to lows mid-August to mid-September. Figure 5, BRW $CO_2$, hourly variability is approximately double that of MLO, with "seasonal" highs in late May, declining quickly to lows in late August. (Details in the Supplemental section).

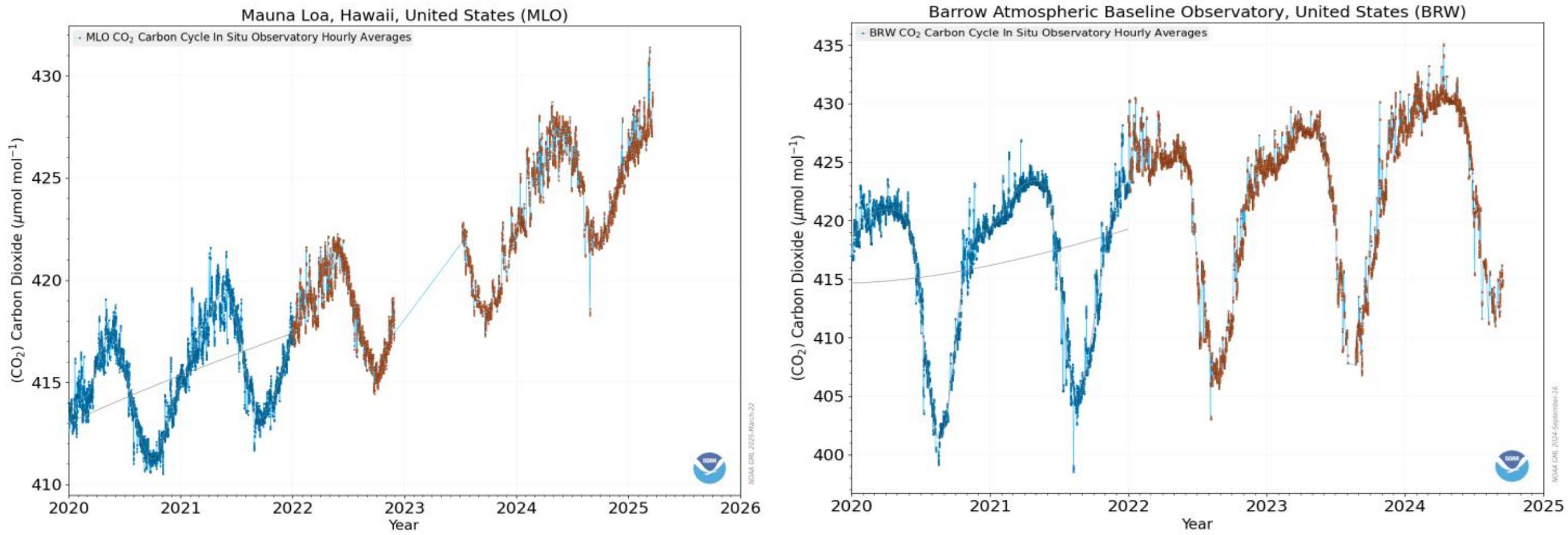

Figure 4. Mauna Loa, Hawaii, (MLO), Hourly $CO_2$ Figure 5. Barrow, Alaska, (BRW), Hourly $CO_2$.
Scale Figure 4: 410 ppb to 430 ppb, 20 ppb. Scale Figure 5 is 400 ppb to 435 ppb, 35 ppb, approx. double Figure 4.

**$CH_4$**. MLO and BRW have large ranges of hourly, daily and seasonal variability of methane, demonstrating their non-linear and highly dynamic nature, Figures 6 and 7. MLO, 2000 to 2016, annual average $CH_4$ concentration increased at an average of 4.32 ppb/yr or .2%. 2016-2025, substantial increase to 12.7 ppb/yr or .6%, three times greater. In 2023 the increase was 22ppb while in 2024 it was 12 ppb/yr. BRW, 2000 to 2016 rate of increase was 4.84 ppb/yr, or .26%. 2016 to 2025, a substantial increase of 8.96 ppb/yr, or .46%, almost double that of MLO.

**Acceleration of the Rate of Increase of Concentration** - Data analysis revealed a substantial acceleration of methane concentration globally and at the two sites. The rate of change of increase, or acceleration, at MLO was 8.6%/yr. and BRW, 4.85%/yr. This acceleration of methane concentration is of concern, due to the much higher global warming effect compared to $CO_2$.

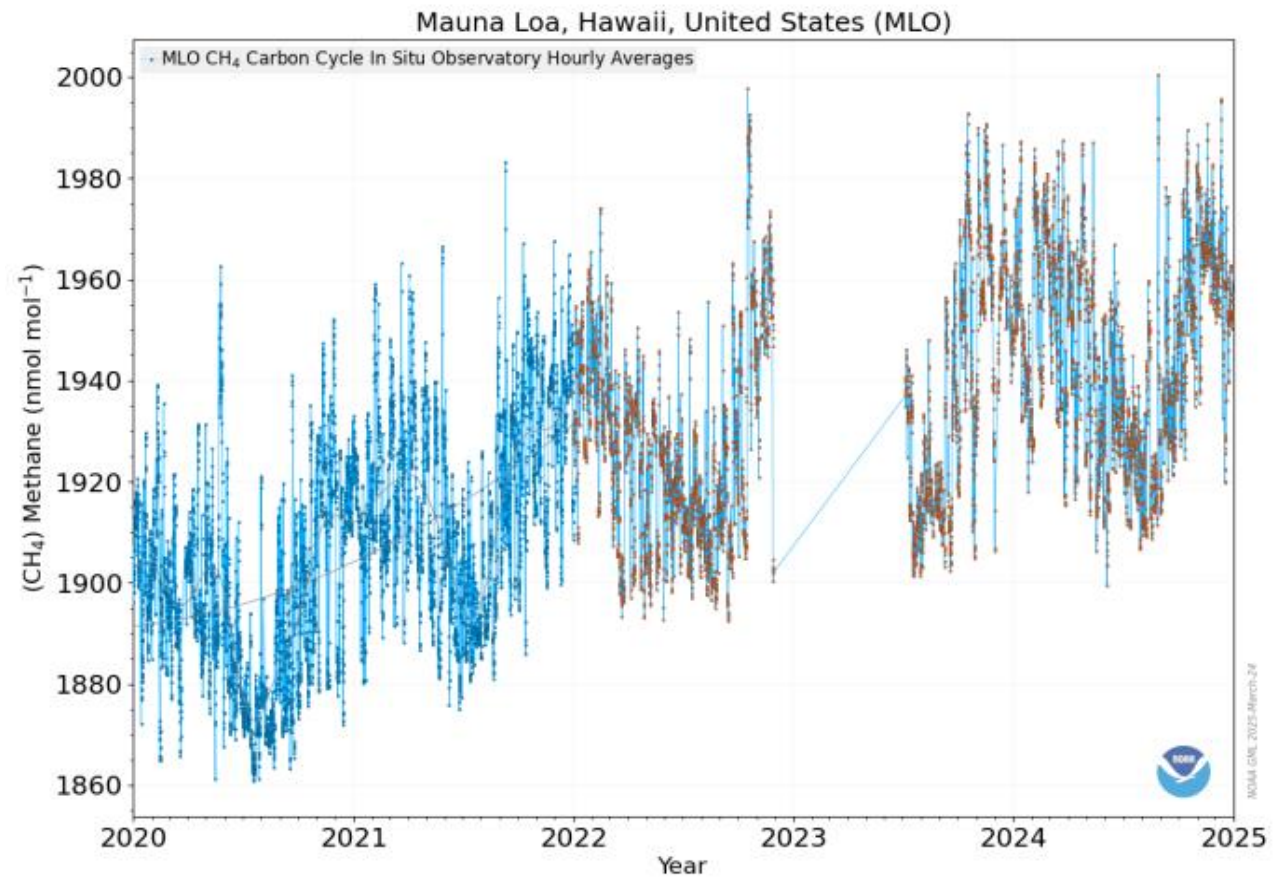

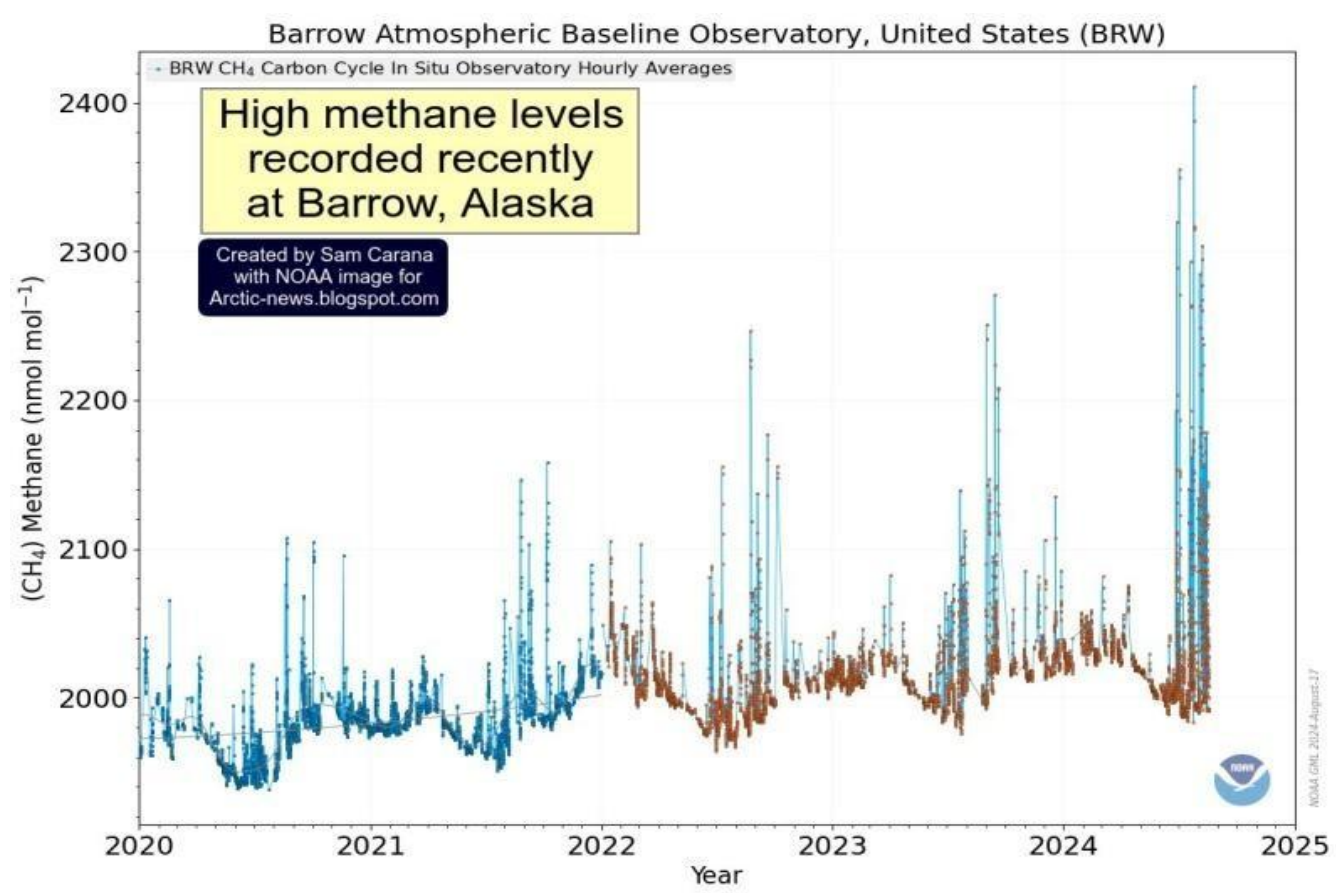

Figure 6. Mauna Loa, HI, hourly $CH_4$ concentration. Figure 7. Barrow, AK, hourly $CH_4$ concentration. Scale of Figure 7 is 3.5 times that of Figure 6. Recent annual range of MLO is 100 ppb; BRW is 400 ppb.

**Emissions Variability and Warming Effect - $CO_2$** - The variability of $CO_2$ was not analyzed in detail, due to the lower variability (Fig's 4 and 5) and the modest warming potential compared to methane.

**$CH_4$** – Methane variability was analyzed: hourly, daily, and seasonally, (details in Supplemental section). Hourly variability is relevant for a highly-dynamic emissions-generating environment, which we contend is the present emissions situation. If hourly variations are substantially greater than weekly, monthly or annual variations, the use of multi-year warming potential such as GWP, or GWP*, over long-term time horizons may greatly misrepresent the atmospheric changes, and concomitant temperatures, and possibly extreme temperature excursions.

Typical seven-day emissions were plotted for MLO and BRW, in January and June, revealing high hourly variations. (see Supplemental Section). Two aspects are important:

- Modelled short-term temperatures using long time horizon data will not be accurate, and
- The dynamics of methane emissions generation may be temperature dependent, a positive, and possibly strong, short-term, feedback loop, with possible micro-climate methane "domes" further accelerating additional methane emissions, on an hourly basis.

June variability, 80 ppb, is substantially greater than January of 25 ppb, by over three times, demonstrating the temperature dependence of methane. These oscillations of methane concentration, prevent attempting to accurately determine the warming effect by modelling. Using a metric such as GWP or GWP* is particularly ineffective, as they use longer time horizons of decades.

**Relevance of Variability of $CH_4$ Emissions** - The variability of methane emissions is particularly relevant for a non-linear, exponential, and temperature-related system, particularly when biological emissions of methane are considered. This applies to waste decomposition, melting perma-frost, and shallow northern-area wet soils and ponds. Biological systems generally respond exponentially to rising temperatures, doubling every 5 to 10 $^{o}$C. Methane generation in anaerobic sediments at 5 $^{o}$C will be minimal but at 40 °C methane production will be more than 100 times greater (Conrad, 2023). In July, 2023 in the Arctic area of Normal Wells, NT, Canada (Lat 65°,16'N), the temperature reached 37.9 °C, 15 ° C above the normal July mean daily maximum. Such temperature extremes are coinciding with substantial increases in methane generation from the Arctic and sub-Arctic.

**Methane "Bursts" of Emissions** - Based on our data analyses, we suggest that some methane emissions, particularly for rapid hourly and daily emissions increases, may be responses to temperature increases, or other factors, possibly in combination, such that in the case of Barrow Alaska, the emissions occur to a large extent, as a series of short hourly and daily "bursts", not a homogenous or constant flow of methane, slowly varying or increasing with time, as assumed in GWP analyses. Such dynamic and variable emissions cannot be modelled using long time-horizons.

## GLOBAL WARMING POLICIES AND IMPACTS

Are GWP, GWP*, GWP* (Improved) and GWP* (Further Improved by a "factor" of 1.13) effective policy tools? Not to date. Emissions continue to increase, as does global warming. The proposed GWP $^{EFF}$ provides a simple, easy-to-use method to compare $CH_4$ reductions for immediate and short-term warming reductions. This is a direct use of Howarth and Balcombe's short-term GWP values, supported by Lund et al., 2000; and identified by Aamaas, (2020). Figure 8, shows that $CH_4$ warming exceeds that of $CO_2$ over 10 years.

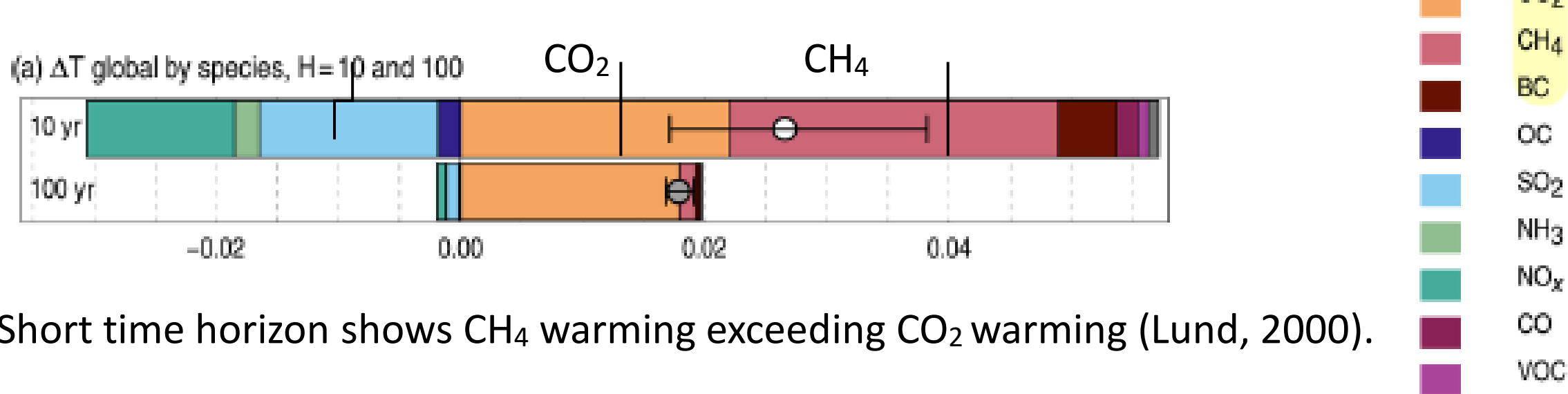

Figure 8. Short time horizon shows $CH_4$ warming exceeding $CO_2$ warming (Lund, 2000).

The under-reporting and misrepresentation of the true impact of methane using $GWP_{100}$ has significant implications for climate policy and mitigation efforts. We conclude that the limitations in the ability of modelling to attribute trends in the sink and source-changing processes are related to uncertainties in the treatment of the model parameters and anthropogenic emissions of methane.

Under-reporting of emissions of methane by exclusion from national inventories such as wildfires, forest waste, waste combustion, and waste treatment, combined with under-reporting of the warming effect has led to distorted and policies that have not been as effective as intended in slowing global temperature increases.

**Underestimating Climate Impacts** - Given the alarming rise in methane levels over the last decade, closer examination of atmospheric methane by climate science and associated modelling is crucial. Accurate and timely climate science is an essential foundation. The growth and pauses observed in the methane records are one of the great challenges in atmospheric chemistry (Heimann, 2020). Simulations in chemical transport modes (Kirschke, 2013) emphasize methane emission estimates and their implementation, by examining trends over the past 30 years in the mixing ratios of methane in the atmosphere.

**Policy Mis-direction** - Focusing solely, or mainly, on $CO_2$ emissions without considering the immediate impact of methane has misdirected and will continue to misdirect resources and efforts to reduce emissions effectively and quickly. Comprehensive strategies targeting methane emissions are crucial for effective climate mitigation. Instead, policymakers in British Columbia, Canada, presently offer electric vehicle subsidies, or heat pump rebates instead of deciding to extinguish wildfires, reduce not eliminate fugitive methane emissions at oil and gas facilities, and waste and treatment locations.

**Missed Opportunities** - Rapid reductions in methane emissions could yield quick climate benefits, slow near-term warming, and provide an emergency buffer, while long-term and slower acting $CO_2$ reduction strategies take effect (Dreyfus, 2022; Shindell, 2024). Extinguishing wildfires quickly reduces emissions and warming and enables carbon sequestration and preserves a valuable economic base for forest communities.

## Discussion

**Temperature and Weather Extremes -** Recent, substantial temperature anomaly increases have not been accurately predicted nor accounted for, nor have substantial temperature variations, or weather extremes and their impacts. Severe weather has affected most parts of the globe with flooding, drought, heat domes, wildfires, ocean die-offs, and severe human health problems: not just extremes, but the extent and frequency of the extremes associated with increasing global emissions.

**Weather Severity –** Viewing typical, nightly news reports, ABC with David Muir during 2025 reveals the shocking pattern of repeated and severe tornados, rainfall and flooding of much of the USA, day after day. Yet, there is no mention of “climate change”.

**GHG Emissions: Concentrations, Rates of Increase, Decay Rate, Variability, Projected Warming -** Among GHGs, methane stands out because of its high radiative efficiency and shorter atmospheric lifespan than carbon dioxide ($CO_2$). Methane emissions in 2023 were estimated at 9.8 Gt CO2-eq (100 year), using IPCC $CO_2$ equivalence of 27.9 . More accurate accounting is required for short-term warming, which would increase the $CO_2$-e to 42.2Gt, whereby methane exceeds warming from 2023 $CO_2$ emissions of 39Gt.

Moreover, the 2024 WMO GHG Bulletin did not consider the short-term GWP of methane, and recommends a focus on reducing carbon dioxide. This illustrates the nature of the problem: under-reported short-term methane warming and the need for its concomitant reduction along with other major greenhouse gases. We highlight these critical gaps and propose avenues to further research and understanding from both mitigation and adaptation aspects in an integrated manner, thereby leading to an effective and unified climate action.

This study identifies both the high variability of emissions as well as the varying decay rate as an important consideration not identified by other researchers, which may help explain recent changes in apparent methane growth rates, annual concentrations, and accelerating global warming.

We have shown that short-term warming effects of annual increases in both $CO_2$ and $CH_4$ are similar, a fact that is not apparent when using a warming time horizon of 100 years, as in the Emissions Gap Report (IPCC, AR6). Overlooking this important fact, that the contribution of methane to global warming is approximately equal to that of $C0_2$, leads to non-optimal policies and mitigation strategies. Not only is the contribution of methane much greater than is generally recognized, but the growth rate increase is double that of $CO_2$, leading to exponential temperature increases. Much of methane's source is biological and is subject to exponential increases with temperature and a vicious feedback loop: coupling temperature increases and methane emissions in a way that $CO_2$ does not.

The exponential rates of growth of methane are shown in Figures 6 and 7. The result of increasing emissions and global warming is that we have already exceeded the UN's limit target of 1.5° C, and that 2° C may be reached by 2030 as Figure 9 (Bryenton, 2025) below shows. A global temperature increase of three degrees C by 2035 is no longer inconceivable.

Calculations, correlations and curve-fitting to recent warming data revealed, not only a second ($x^2$) or third order ($x^3$) exponential, but that a fourth order ($x^4$) was the best fit to recent data, to project future temperature possibilities. If warming increases at recent rates, 2° C may occur by 2030.

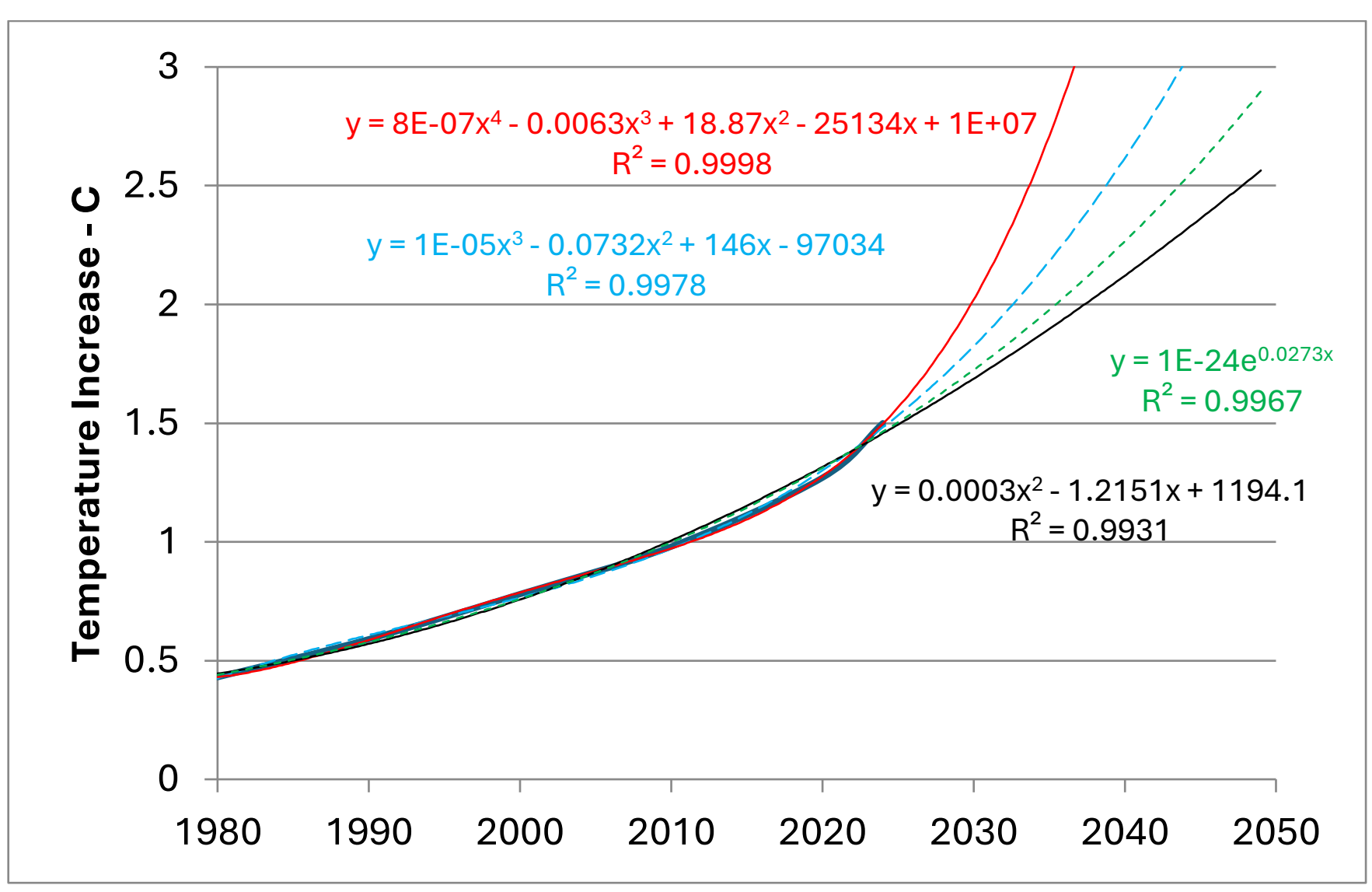

Figure 9. Temperature Increase by Year. Plausible temperatures after 2024; 2°C by 2030. Bryenton, 2025

Recognizing methane's contribution to warming is critically important to reduce fugitive emissions from fossil fuel extraction, waste and wildfires. Wildfires are not directly "anthropogenic" and thus not included in published federal emission inventories. They are "not included" but published in British Columbia's provincial inventories and have recently exceeded all other GHG emissions. Wildfires are more frequent and "hotter" (Vaillant, 2024), due to lower humidity, larger, wind-driven fires with increased radiant energy, spreading faster, with greater destruction and additional methane emissions. These emissions are directly influenced by humans' response time, effort and available fire-fighting equipment resources.

**Temperature Prediction Failures Under Steady-State Assumptions -** The systematic underestimation of methane's climate impact, through stead-state assumptions, may partially explain why recent global temperatures have exceeded model predictions. When climate models incorrectly assume immediate methane equilibrium, they fail to capture the substantial warming momentum created by methane accumulation over the past several decades.

Our recalculation using $GWP^{EFF}$ suggests that cumulative methane forcing may be from four to seven times greater than estimated using steady-state assumptions. This "missing warming" from accumulated methane could account for a significant portion of the temperature anomalies observed in 2023-2024, when global temperatures exceeded 1.5$^{o}$C above pre-industrial levels earlier than most models predicted.

The prevalence of this modelling error in climate literature represents a systemic failure to properly represent short-lived greenhouse gas dynamics, with profound implications for both climate prediction and policy effectiveness.

## POLICY IMPLICATIONS AND CASE STUDIES

Addressing methane emissions from an immediate GWP perspective, $GWP^{EFF}$, necessitates targeting major methane sources such as oil and gas production, electricity from fossil fuels, hydropower, agriculture, waste management, and residential methane gas use for heating and cooking. MethaneSAT (2024) revealed that oil and gas industry emissions were four times greater than those of the US Environmental Protection Agency, and eight times greater than estimates made by industry. In addition to this under-reporting, which uses IPCC $CO_2$–e of 27.9, when converted to a short-term value of 120, further increases the under-reporting by an additional factor of 4.3. Overall, this results in more than 32 times the "reported" methane emissions. We highlight three cases that support our findings and call for urgent methane reductions.

**Wildfires and Methane Emissions** - We propose that reducing wildfire emissions is an unrecognized and effective strategy for addressing global warming. Following Canada's National Inventory Report (NIR, 2022, Table A6.61), approximately 98% of wildfire emissions are $CO_2$, with 1/90$^{th}$ or 1.11 % methane. We use this study's $GWP^{EFF}$ and "warming units", calculated using Eq (1), which uses an immediate GWP-GTP of 120 for methane. With methane emission of 1.11% of emissions, this results in 0.011 x GWP of 120 = 1.333 "warming units". $CO_2$ at 98% × 1 GWP = .98 "Warming units". Thus, the methane warming effect from wildfires is 1.36 times greater than that of $CO_2$ in the short term. We suggest that this methane warming from wildfires, not reported by other researchers, contributes substantially to recent temperature extremes. This information will assist policy makers in directing additional resources to reducing methane-related emissions which may be overlooked, with many focussing instead on reducing $CO_2$ emissions. Canadian wildfires in 2023, unprecedented, burned an area nearly the size of Ireland, releasing approximately three billion tons of $CO_2$ which is equivalent to four times the carbon emissions of the global aviation sector (WRI, 2024). Byrne et al. (2024), determined 647 Tg or 647 Mt of carbon was emitted, equivalent to 2.37 Gt of $CO_2$ . Adding methane emissions at 1.11% increases the warming effect by an additional 3.16 Gt of $CO_2$-e. When compared to total global emissions, we note that Canada's fires increased emissions by over 3 %.

Canadian fires accounted for approximately 23% of global wildfire carbon emissions in 2023 (Copernicus, Global News, 2023). We note that as a result of the wildfires, the global warming contribution of methane thus becomes fully 14% of total anthropogenic GHG emissions.

Not only were fire-related emissions substantial, including methane, but also the loss of future $CO_2$ sequestration must be considered: another essential avenue of investigation to determine the cumulative effect of wildfires on global warming.

**British Columbia: "Reported", Anthropogenic vs. "Actual" Emissions - I**n 2017, British Columbia, Canada, "reported" total GHG emissions of 64.4 Mt, $CO_2$e (BC Government, 2021). Wildfire emissions of 235 Mt. $CO_2$-e (using $GWP_{100}$ of 25) were almost 4 times greater than all "reported" emissions. By gas, including $CO_2$ and $CH_4$ , using a $GWP_1$ (one year) of 120, the "reported" anthropogenic methane emissions of 8.8 Mt were dwarfed by the recalculated 125 Mt of methane actually emitted into the atmosphere, which included wildfires. This further demonstrates the general under-reporting of the warming effect of methane, across most economic sectors and by most emissions reporting agencies.

**Liquefied Methane or "Natural" Gas (LNG) Sector** - The LNG sector is another significant source of methane emissions. Methane leakage during the extraction, processing, and transportation of methane gas substantially impacts the environment. Methane losses could be as high as 3-4% of total methane production (IEA, 2021). This sector's emissions underscore the importance of implementing stricter controls and adopting new technologies to minimize leaks and reduce methane emissions. Further-more, when deciding whether to vent or flare methane, the decision is facilitated by examining the warming potentials: $CO_2$ from flaring combustion at a GWP of 1, or vented methane at a $GWP_1$ of 120: flaring is the logical option.

## Conclusions

Methane's short-term climate impact is profound, and drastically underestimated by the traditional GWP metrics. We propose the adoption of realistic short-term GWP values, using $GWP^{EFF}$ to enhance our understanding of the role of methane in global warming and drive more effective policy measures. Immediate action on methane emissions is crucial for achieving significant near-term climate benefits, to address the "climate emergency", while mitigating long-term climate risks.

Not only were Canadian and global wildfire emissions substantial, including methane, as above, but also the loss of future $CO_2$ sequestration must be considered: another essential avenue of investigation to determine the cumulative effect of wildfires on global warming.

**RECOMMENDATIONS: CHALLENGES AND OPPORTUNITIES**

Adopting a very short-term GWP, such as $GWP^{EFF}$ as a more immediate GWP, faces resistance from traditional emissions accounting frameworks and scientific conservatism. It was in 1990, the IPCC (Climate Change) and again in 2014 when Howarth's immediate GWP of 120, a 10-year GWP of 105, and a 20-year GWP of 86 were introduced. Yet 10 years later, the 2024 Emissions Gap Report continued to use a $GWP_{100}$ value of 28.5. The latest WMO GHG report (2024) also does not consider the short-term impacts of methane, but recommends that carbon dioxide be reduced as a focus, which we partly agree with, however, methane and its reduction must also be placed as high priority.

The urgency of addressing the "climate emergency" demands innovative approaches. We call for revisiting and researching in-depth, the immediate warming impacts of methane by incorporating the highly variable and short-term aspects of methane into all calculations. Using short-term GWP, or $GWP_1$ and including the use of "warming units" or $GWP^{EFF}$ is a promising avenue to pursue.

In addition to refining GWP metrics, another crucial area of research is revisiting climate science and associated modeling to better simulate the feedback mechanisms and non-linear processes that characterize the sources and interactions of methane with the atmosphere. The recalculated short-term warming potential of methane using a GWP of 120 highlights its significant and immediate impact on global warming, emphasizing the need for integrated climate strategies and IPCC re-evaluation of short-term GWP values. Combining rapid climate mitigation with adaptation measures can potentially provide swift climate benefits, aligning with the call for unified adaptation and mitigation approaches to enhance policy coherence and effectiveness (Howarth & Robinson, 2024).

**ROLES AND RESPONSIBILITIES**

Researchers, the public and policy makers all have responsibilities:

1. Increasing emissions monitoring and analyses: more frequent, and less delay (presently 1 year).

2. Urgent development and implementation of policies committing to a planned phase out of existing and proposed fossil projects, prioritized by level of emissions, including replacing fossil projects with renewable energy sources.
3. Immediate methane emissions reductions of “reported” and “not included” sources, which will also reduce emissions from “natural” sources by reducing atmospheric temperatures.
4. Increasing funding for climate research and emissions-reduction programs, including government and private, with penalties for target failures.
5. Improved and expanded communications among researchers, policymakers, and the public. Scientists must directly communicate, formally, with policymakers, elected officials, and the public, while policymakers must seek input from scientists and the public, and the public must interact with both scientists and policy-makers to seek immediate consensus and agree to mitigation strategies and implementation.
6. Integrate climate mitigation with adaptation measures to unify effective climate action.

COP 30 presents an ideal opportunity for obtaining consensus on improving the analyses of methane’s critical role in global warming and fixing the “Broken Record”, and eliminating “Hot Air“ to facilitate emissions reductions. By embracing these innovative approaches, we can better align our scientific understanding with the urgent demands of the escalating climate emergency. This will ensure that scientific advancements contribute directly to effective, immediate and integrated climate adaptation and mitigation strategies.

## Acknowledgements

The authors would like to acknowledge the work of Sam Carana, who provided access to Figure 4, an hourly methane plot of in situ data, provided by NOAA, from Barrow, Alaska. We also acknowledge the helpful review of the research and paper by Borgar Aamass, CICERO, Center for International Climate Research, Oslo, Norway. We acknowledge that there was no financial support for this work; it was entirely undertaken by independent researchers in the interest of expanding climate-change knowledge and emissions reductions.

## Competing Interests

The authors declare no competing interests.

# Open Research

Data Availability Statement. Supplemental Data and Detailed Analyses are available. Supplement to the paper provides additional data, analyses and discussion.

Figure 1. Open Access. CO2 equivalence over different time horizons. GWP and GTP are equal for 10 years (Balcombe et al, 2018). *Environmental Science: Processes & Impacts, 20* (1323-1339). ). DOI: 10.1039/c8em00414e https://doi.org/10.1039/C8EM00414E

Figure 2. Open Access. Over a ten year horizon, (left panel), warming from methane emissions (second, brown bar) exceeds that from $CO_2$ emissions, (first, red bar)( Howarth et al, 2014). A bridge to nowhere: Methane emissions and the greenhouse gas footprint of natural gas. *Energy Science & Engineering, 2*(2), 47-60. https://scijournals.onlinelibrary.wiley.com/doi/10.1002/ese3.35

Figure 3. Sam Carana, 2024. Methane emission in Arctic regions greatly exceed past concentrations. *Arctic-news.blogspot.com, September 4, 2024.* Data Sourced from NOAA, Barrow, AK, Data Visualization, Carbon Cycle Gases, Time Series, CH4, In Situ, Data Frequency – Hourly Averages, Start Year – 2020, End year – 2025, Submit.
https://gml.noaa.gov/dv/iadv/graph.php?code=BRW&program=ccgg&type=ts

Figures 4, 5 and 6. Data Sourced from NOAA. Mauna Loa, HI and Barrow, AK. Data Visualization, Carbon Cycle Gases, Time Series, $CO_2$ and $CH_4$, In Situ, Data Frequency – Hourly Averages, Start Year – 2020, End year – 2025, Submit.

Emissions Gap Report Data: Use of GWP $CO_2$ e 25, IPCC AR5 and 6: *Sources:* Crippa *et al.* (2023) for GHG emissions; Friedlingstein *et al.* (2022) for bookkeeping LULUCF $CO_2$; Grassi *et al.* (2023) for inventory-based LULUCF $CO_2$. https://doi.org/10.59117/20.500.11822/43922

Wildfire emissions data.
Canada, Methane amount of 1/90th or 1.111 percentage from:
https://www.google.ca/url?sa=t&source=web&rct=j&opi=89978449&url=https://data-donnees.az.ec.gc.ca/api/file%3Fpath%3D%252Fsubstances%252Fmonitor%252Fcanada-s-official-greenhouse-gas-inventory%252FD-Emission-Factors%252FEN_Annex6_Emission_Factors_Tables.xlsx&ved=2ahUKEwja__fPzNmIAxXqCTQIHWaGLR0QFnoECBUQAQ&usg=AOvVaw3RdxtmFctDDHu103jIpXVD

Comparison with global aviation sector from: World Resources Institute. (2024). Article. Canada's 2023 Forest Fires Caused Major Climate Impact. https://www.wri.org/insights/canada-wildfire-emissions

Emitted carbon from Canada's fires, by Byrne et al from: Byrne, B., J. Liu, K. Bowman, M. P-Campbell, A. Chaterjee, S Pandey, K Miyazaki, G. vander Werf, D Unch, P. Wennberg, C. Roel, S Sinha, Carbon emissions from the 2023 Canadian wildfires, *Nature,* Article, August 2024.. DOI:10.1038/s41586-024-07878-z

Canada’s 2023 fires were 23% of global total from:  Global News. (2023). Canada produced 23% of global wildfire carbon emissions for 2023. https://globalnews.ca/news/10167079/canada-wildfire-carbon-emissions-2023/

British Columbia wildfire emission data from: Province of British Columbia Greenhouse Gas Emissions: https://www2.gov.bc.ca/assets/gov/environment/climate-change/data/provincialinventory/2021/provincial_inventory_of_greenhouse_gas_emissions_1990-2021.xlsx

## Funding Statement

The study was undertaken by independent researchers in Canada and Pakistan without external funding or support, in order to provide greater access to climate change data and methane analysis under Open Access.

## Ethical Compliance

There were no procedures performed in studies involving human or animal participants.

# Supplemental Data and Analyses

**DETAILED, RELEVANT CLIMATE SCIENCE CONCEPTS FOR POLICY MAKERS**

**Global Temperatures** - Data were obtained for global temperatures and variability, with the last four years the warmest on record, and with 2024 exceeding the IPCC target of 1.5 °C. $CO_2$ and $CH_4$ emission data were accessed for two locations, one in the Arctic at Barrow (BRW), Alaska, the other at Mauna Loa (MLO), on Maui, Hawaii. This was to assess variability on hourly, daily, seasonally and annual bases including an examination of non-linearity: also hourly, daily and annually. The two sites incorporated high latitude and lower latitude regions, both in areas with low populations and anthropogenic emission sources

**Temperature Extremes –** 2022 was 0.86° C above the 20th century average of 13.9 °C; 2023 at 1.18 °C; 2024 at 1.55 °C (WMO). Copernicus reported 2024 as 1.6 °C. "Each of the past 10 years was one of the 10 warmest years on record", while a "new record high for daily global average temperature… 22 July 2024 at 17.16 °C" (WMO), the basis for Hausfather's comment, "a bit warmer …than anticipated".

**Warming and Time Horizon** - Emissions data and $CO_2$ "equivalency" values ($CO_2$-e) for $CO_2$ and $CH_4$ were obtained and analyzed. Literature review provided different "warming factors" for methane, depending upon the time horizon chosen which we applied to published emission reports, as "corrections". Most time horizons for assessing global warming, as used by IPCC in determining and reporting emissions, are based on 100 years, while some researchers use up to 500 years. (Fuglestvedt, 2000). These time horizons are not useful for short or very short lifetime gases or their impacts, such as immediate to 5 or 10 years. Thus, emissions reporting for methane, such as domestic waste, decomposition of forest harvesting waste, agriculture, or oil and gas extraction, under-estimate the warming impact over short time horizons (Howarth, 2014; Balcombe,2018). Immediate warming from methane can be as much as 120 times greater than $CO_2$ (Howarth, 2014; Balcombe, 2018), thus the 100-year value substantially under-reports the actual warming effect by over four times.

The Emissions Gap Report (EGR, 2024 for anthropogenic $CH_4$ emissions of 9.8 Gt $CO_2$ –e used a GWP of 28. When adjusted for the short-term effect, this becomes 42 Gt, an increase of almost 4 times. Methane's warming effect diminishes rapidly with time which was assessed for annual replacement quantities of 8%/yr and compared to the annual increase of 0.5%, further described under "Decay Rate".

**Global Greenhouse Gas Concentrations and Rates of Increase** - The principal global warming GHGs, $CO_2$ and $CH_4$ have atmospheric concentrations of approximately 426 ppm (parts per million) for $CO_2$ and **1923** ppb (parts per billion) or 1.92 ppm, one 200$^{th}$ as much for $CH_4$. $CO_2$ is increasing at 2.4 ppm or 0.5% per year. Methane concentration increased at a range of 9 to 17 ppb in 2021, almost 1% per year, double that of $CO_2$. The 2023 increase was lower at 8.51 ppb. $CO_2$ has a long lifetime, decaying very slowly, while methane decays approximately 8.4% per year. We identified the high rate of decay as a major factor in GHG accounting. This rate of change was not reported by other researchers nor was the increase in the rate of change, discussed below.

$CO_2$ Emissions - Annual global fossil $CO_2$ emissions, as per EGR, 2024, Table 2.1 were 39.1 Gt.
$CH_4$ Emissions – Annual global total methane emissions are from 575 Tg (Mt) (calculated bottom-up) to 669 Tg (Mt) calculated top-down (Global Methane Budget, 2025). Averaging these provides a mid-range value of 663 Tg (Mt). Anthropogenic methane emissions are from to 331 Tg(Mt) to 343 Tg(Mt), with a mid-range of 337 Tg (Mt), or 54% of total. This agrees closely with the Emissions Gap Report, Table 2.1 value of 350 Mt, or 9.8Gt $CO_2$-e using a conversion of 28 for a 100-year warming factor. When corrected for immediate, short-term GWP of 120, the effective warming increases to 42 Gt $CO_2$ –e, exceeding warming from $CO_2$.

**Emissions Variability** – Literature review did not identify that with highly variable emissions, the use of long-term warming factors such as GWP or GWP* over an extended time horizon, such as 100 years does not accurately assess the timing or magnitude of warming. This is particularly relevant for methane with high emissions variability, both hourly and seasonally, and with rapid decay involved.

**Methane Decay and Replacement** - The decay rate of methane is approximately 8.4% per year, with a half-life of 8.4 years and a perturbation life of 12 years (IPCC AR5 and AR6). Nisbet et al. (2023) calculated the ratio of $CH_4$ emission rate to concentration rise to be 2.77 Mt of methane per ppb. This has profound implications for global annual emissions-reporting as it is unclear if methane decay and replacement is reported in various emissions reporting publications. Total anthropogenic and natural global annual methane emissions are up to 669 Mt in a top-down analysis (Global Methane Budget, 2025) of which 561 Mt is decay and replacement methane, almost 84%.

**Policy Implications of Under-reported Methane Emissions** - Existing policies to reduce GHG emissions using under-reported emissions analyses cannot provide optimum solutions. The largest sources of methane may not receive the largest reduction efforts: oil, gas and coal mining activities, anthropogenic waste, reducing wildfire-related methane emissions; or agricultural sources from ruminants and rice production.

Policies and efforts to reduce emissions from $CO_2$ sources may overlook greater benefits of methane reduction efforts. Increased methane emissions from "natural" sources due to warming, can potentially be reduced indirectly as anthropogenic sources are reduced and lead to reduced warming. Failure to capture the urgency of methane mitigation for near-term temperature stabilization can result. Climate, health, and financial benefits from methane reduction policies are not achieved.

**Broader Scientific Implications -** Policy ineffectiveness: Current strategies fail to address the most immediate warming drivers. Climate emergency acceleration: Accumulated methane creates warming momentum not captured in steady-state models.

**Global Temperature Modeling** – Measured temperature increases continue to exceed climate scientists' predictions. Measured temperatures are to the upper range of the anticipated temperatures, 1.55 ° C vs predicted 1.3 ° C, a difference of 20%, Figure S1.

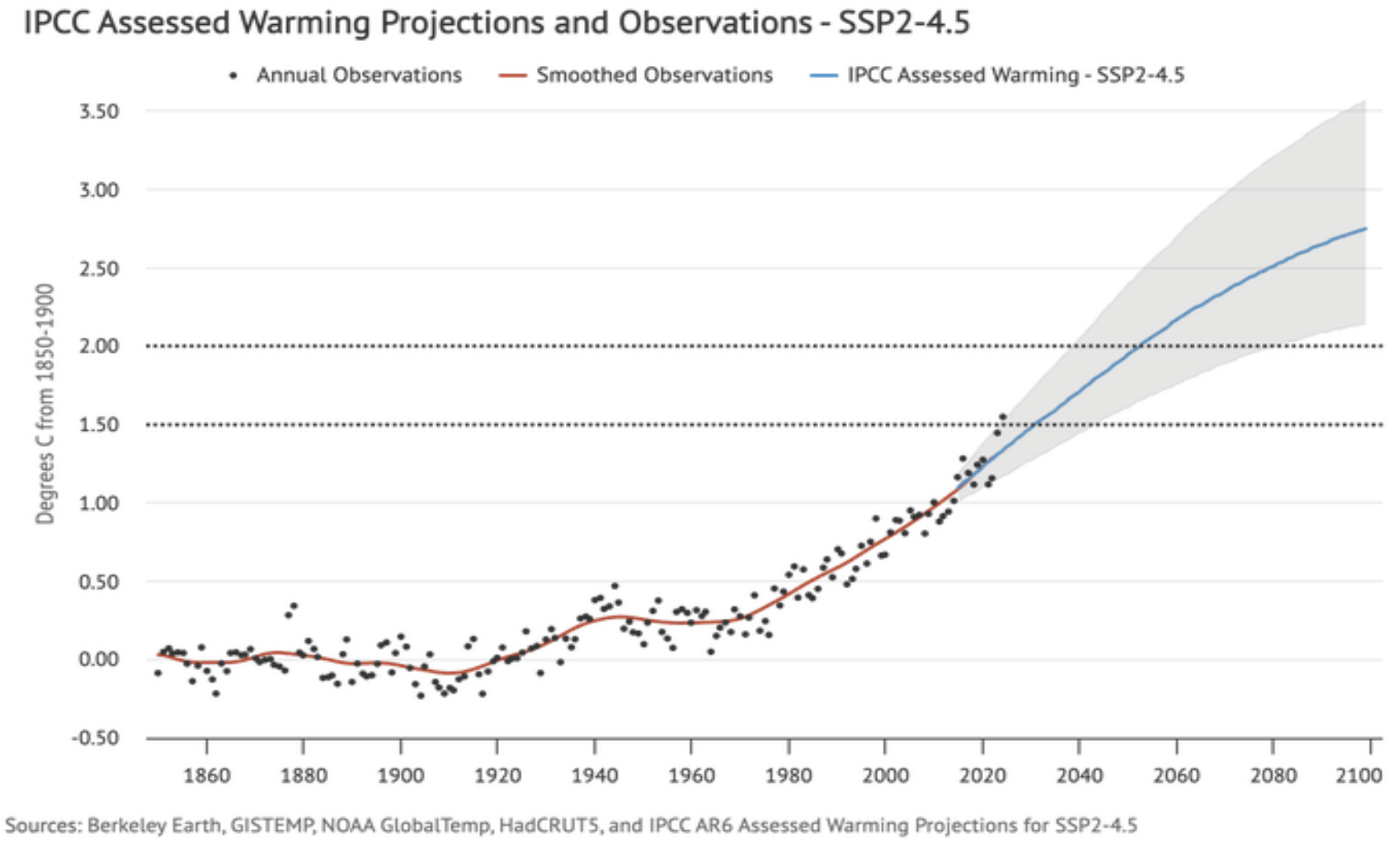

Figure S1. Temperature Modeling. Measured temperatures exceed modeled temperatures (Berkeley Earth).

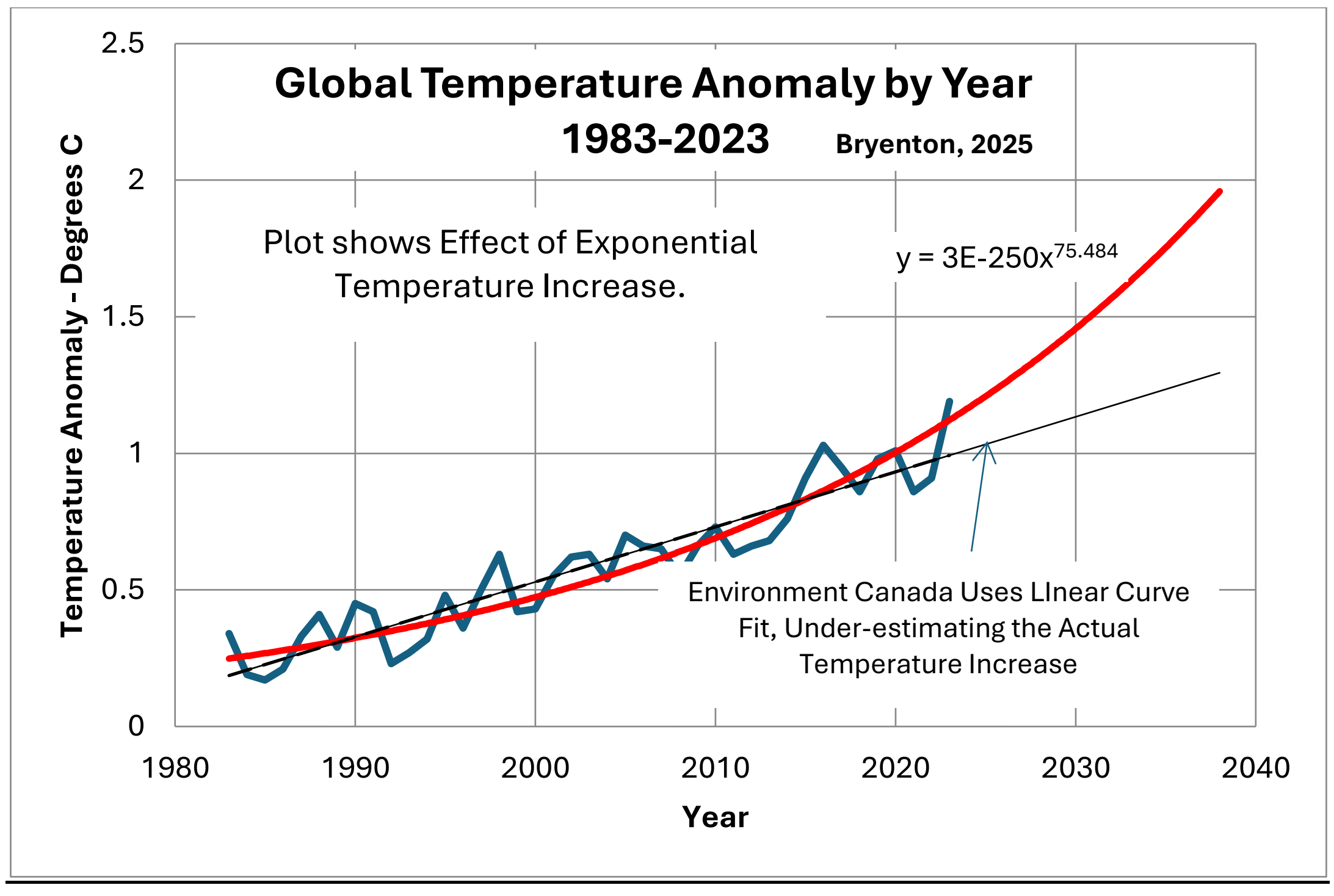

Figure S2. Non-Linear, Exponential Global Temperature Increase. Based upon an exponential increase it is plausible that global temperatures could increase to 2° C by 2040. Bryenton, 2024

**<u>Temperature, Methane and Carbon Dioxide Correlation</u>**. – Temperature Increase is exponential, (red line on Figure S2, above). $CH_4$ increase is exponential, (orange line), Figure 3 below; $CO_2$ increase is linear (black line). Temperature and methane - both exponential: methane is driving temperature.

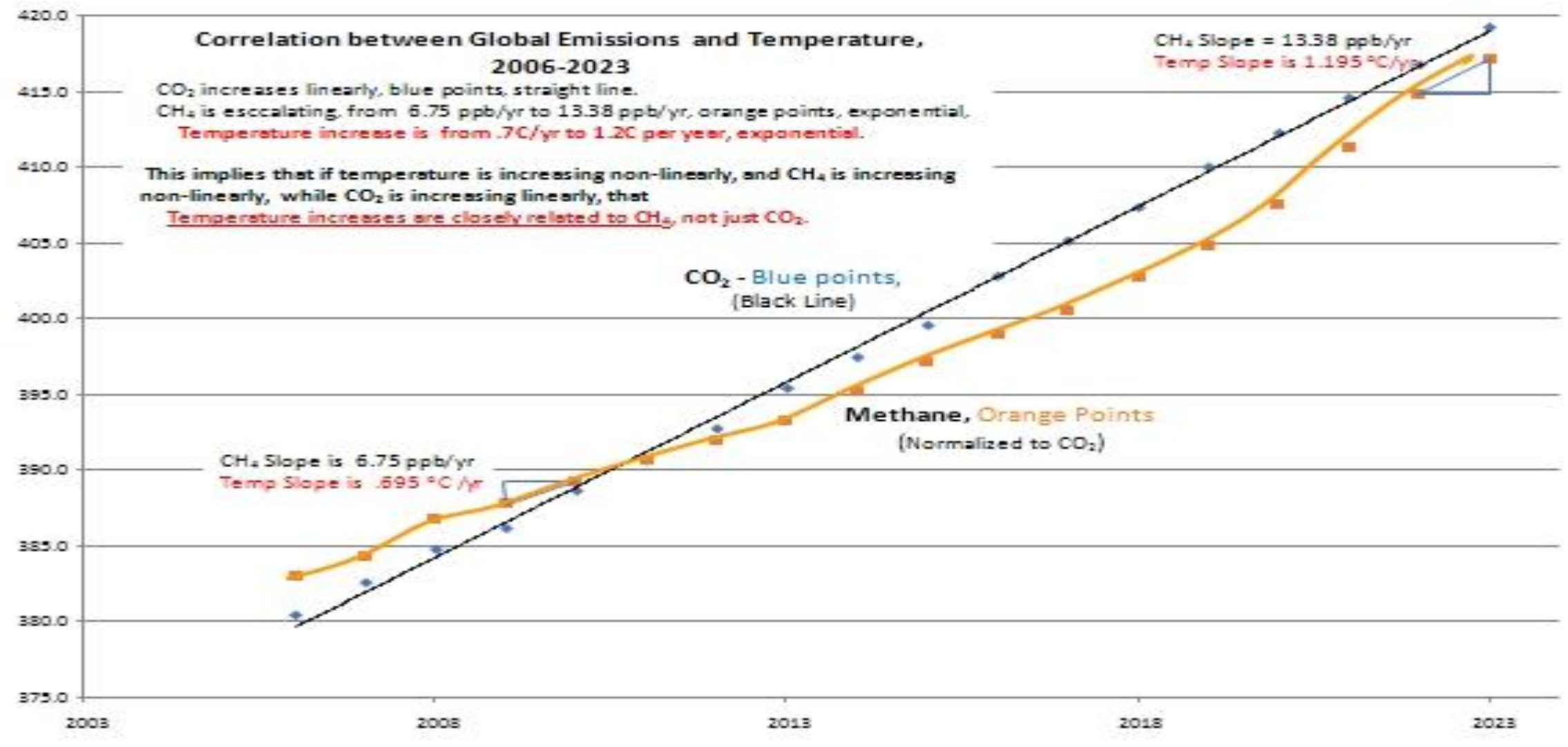

Figure S3. Plot of Exponential Methane Increases, and Linear Carbon Dioxide Increase, 2006-2023, compare to linear increase of $CO_2$ from Figure S2 above. (Bryenton, 2025).

**Variability: $CO_2$ and $CH_4$ at Mauna Loa, HI and Barrow, AK: Hourly, Daily and Seven Day (Weekly).**

<u>$CO_2$ Emissions and Variability</u> - The variability of $CO_2$ is less important for warming compared to methane, due to methane's much greater warming effect over short-term time horizons: daily, hourly, monthly and seasonally. Figure S4 for MLO shows $CO_2$ emissions have considerable variability: hourly, daily, monthly and seasonally, with highs late spring, May, declining quickly to lows mid-August to mid-September. Figure S5 for BRW $CO_2$, shows hourly variability of approximately double that of MLO, with "seasonal" highs in late May, declining quickly to lows in late August.

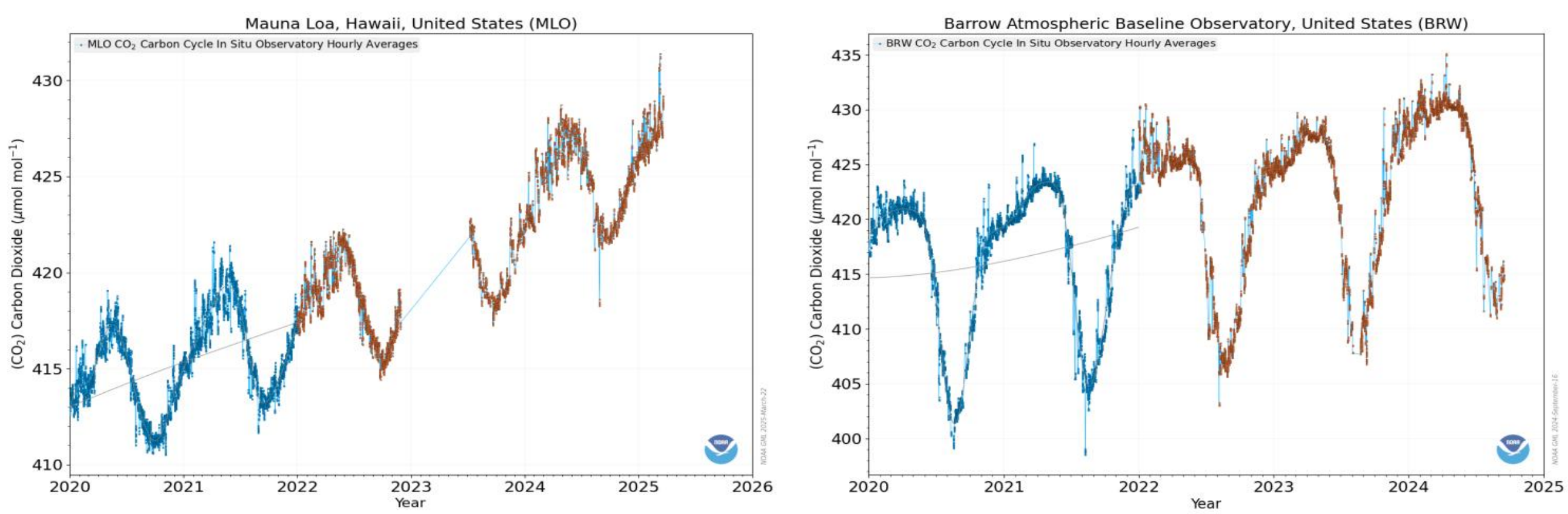

Figure S4. Mauna Loa, Hawaii, (MLO), Hourly $CO_2$ Scale Figure S4: 410 to 430 ppb, 20 ppb.

Figure S5. Barrow, Alaska, (BRW), Hourly $CO_2$. Scale: Figure S5: 400 to 435 ppb, 35 ppb, approx. double.

|  |  |
|---|---|
| **MLO** - Average $CO_2$ concentrations (light grey line) are slightly lower than Barrow, AK, Figure 4. Annual variability is 10ppm, 2.3%. | **BRW -** $CO_2$ is more variable than MLO. Annual variability increases from 22 ppm or 5.96% in 2000, to 33.6 ppm (or 7.82%) in 2024. Increase of 11.6 ppm, or 53%. |
| **Emissions range** - 2020 to 2025, total variation was 18.5 ppm, or 4.4% of the "average" of 420 ppm. | **Emissions range** – 2020 to 2025, total variation was 39 ppm or 9.3% of "average" of 418 ppm; **double MLO variability** |
| **Annual emission variability increases** from 8 ppm, 2000, 2.1%, to 12 ppm, or 2.9%, over 14 year period, a ppm increase of 50%. | **Annual variability compared to MLO is 2.7 times greater**. Annual variation range: 26 ppm or 6.1%. 2.48 times greater than MLO. |
| In 2024, the **annual variation range** (NOAA charted hourly data) was 8.65 ppm or 2% (ignoring one extreme low event of 428 ppm). | In 2024, the **annual variation range was 3 times greater than MLO;** 24 ppm or 5.7% % (ignoring an extreme high event of 437 ppm, late December). |
| **Annual average rate of change**: from 2000 to 2016 it was 2.3 ppm/yr, while from 2016 to 2024 the rate of change was 2.5 ppm/yr, an increase of 8.6%. | **Annual average rate of change**: was 2.07 ppm/yr, while from 2014 to 2024 the rate of change was 2.57 ppm/yr, a substantial 24% increase, almost three times MLO. |
| **Hourly** variation, 2024, ranged from increases of 5.86 ppm or 1.4 % to decreases of 4.4 ppm or 1.7 %. | **Hourly** variation, 2024, ranged from increases of 4.92 ppm or 1.1 %, to decreases of 4.41 ppm or 1.0 %. |
| **Seasonally**, MLO 2024, peak emissions were from April through June, of 436 ppm, with rapid declines to September lows of 417 ppm. Lows are late summer to early fall. | **Seasonally,** BRW 2024 peaks were in September and December 438 and 437 ppm, with lows July and August of 405 ppm ,slightly earlier than MLO lows. |

$CH_4$ Emissions and Variability - MLO and BRW have large ranges of hourly, daily and seasonal variability of methane, demonstrating their non-linear and highly dynamic nature, shown in Figures S6 and S7. GWP and GWP* are able to accurately represent emissions warming over time frames of 20 years or greater, due to assumed relatively stable emissions rates. However, determining a short-term "warming" potential using GWP or GWP* may involve considerable temporal errors. Hourly calculations of warming would be appropriate for modelling temperatures, however, this would far exceed the capabilities of most policy makers.

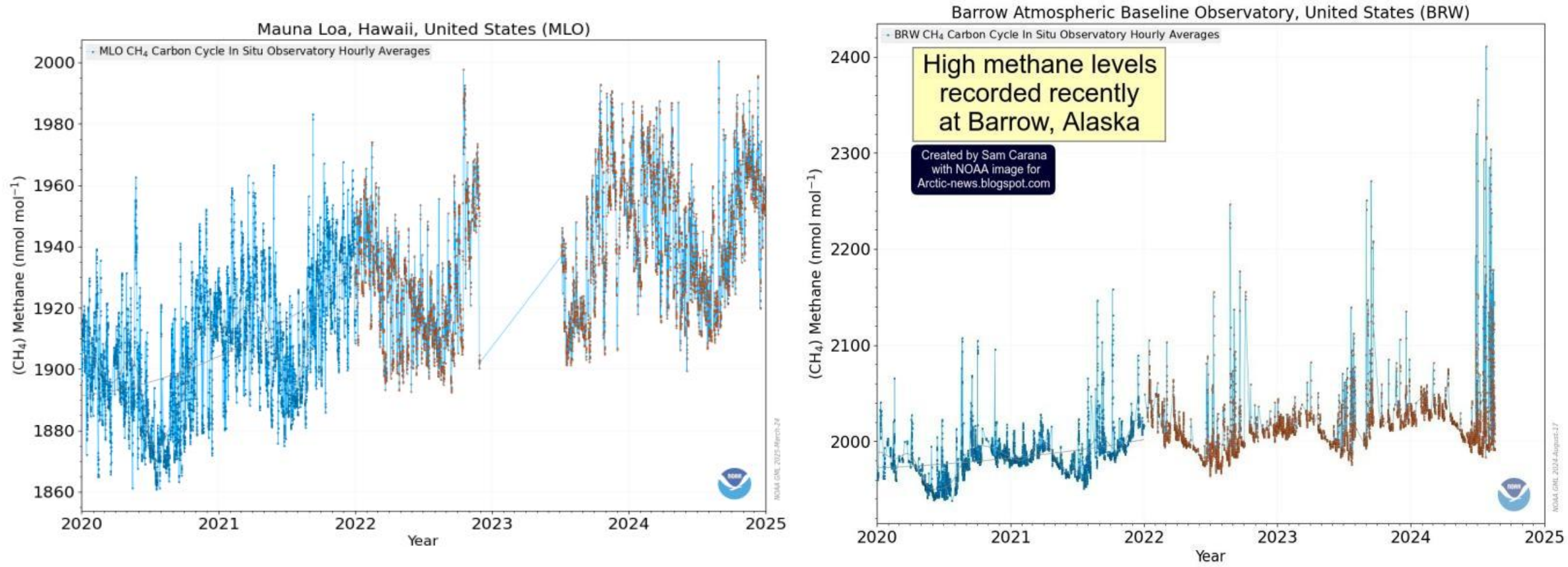

Figure S6. MLO, HI, hourly methane concentrations. Figure S7. BRW, AK, hourly $CH_4$ concentrations.
Scale of Figure S7, BRW variability, is 3.5 times that of Figure S6, MLO.

Recent annual range of MLO is 100 ppb; BRW is 400 ppb.

| **MLO** concentrations are lower than BRW, partly due to high elevation 3,400m or 11, 000 ft. and distance to source. | **BRW**, recent emissions range greatly exceeds past variability. (Carana, 2024). |
|---|---|
| **Multi-year, hourly concentration variability** ranges from 90 to 104 ppb, or 5%, over the 5 year period, 2020 to 2025 Barrow is more than double that of Mauna Loa**.** | **Multi-year, hourly variations** of $CH_4$ concentrations are **highly variable.** Hourly increases of 279 ppb or 11.5%, and decreases of 238 ppb or -11% over the 5 year period. |
| Total annual range in 2020= 102 ppb or 5.3% to 2024=98 ppb or 4.8%. Little change. | Total annual range in 2020 = 170 ppb, or 8.4%, to 2024 =378 ppb, or 17%. **A doubling of emissions range in four years. 2024, emissions range was almost 4 times that of MLO.** |
| Over 5 years**, 2020 to 2024, total range** was 149 ppb, or **7.75**% of "average" of 1922 ppb. | Over 5 years, **total range** was 445 ppb, **or 20.6**% of "average" of 2163 ppb. T**otal range almost 3 times MLO.** |
| **Accelerating growth rate** of peak concentrations from 2020 to 2024 increased 38 ppb, or less than 2% in 5 years. | **Accelerating growth rate** of peak concentrations from 2020 to 2024 **increased 275 ppb, or 13% in 5 years. Over 7 times the increase of peaks in MLO.** |
| **Hourly** variations during **2024** ranged from 30 ppb **or 1.5%** to – 26 ppb or 1.17% (NOAA hourly data) | **Hourly** variations during **2024** ranged from 193 ppb or 8.5 %, to - 135 ppb **or 6.7 %.** |
| **Extreme, peak values.** There was no substantial increase of peaks nor decrease of lows over the 2020 to 2025 period. | E**xtreme, "peak" values** were noted, in 2023, of over +278.8 ppb and – 237 to 231ppb, however substantial variations in nearby hourly values indicated possible data errors. |
| **Seasonally**, 2024, peak concentrations were scattered, from January to early May, and from October through early December, with the maximum in mid-October of 2001 ppb. | **Seasonally**, 2024, peak concentrations were scattered, from May, to September, and late November through December, with a plausible maximum, early September, of |

| MLO lows were late May through mid-August, with the lowest mid-May of 1903 ppb. | 2285 ppb. A questionable maximum was mid-September of 2385 ppb. Lows were mid June, mid-July and mid-August, with the lowest plausible value of 1990. There was an extreme low at the end of June of of 1960 ppb. The **plausible variation in 2024 was 294.5 ppb,** or 13.8%. of concentration.<br><br>**Extreme hourly variations** occurred ranging from over +250 ppb to -237 ppb in late June; possible to data integrity concerns, and errors.<br><br>A second plausible explanation by these researchers is that at Barrow, AK, **rapid methane increases may be due to "bursts",** often followed by rapid dissipation, rather than steady flows of methane. Additional research is required. |
|---|---|
| From 2000 to 2006 **the annual average rate of change** was 1.7 ppb/yr and from 2020 to 2025, 14 ppb/yr. This is an **increase approaching 10 times**, clearly demonstrating the non-linearity and exponential increase of methane | From 2000 to 2006 **the annual average rate of change** was 1.67 ppb/yr and from 2020 to 2025, 13.4 ppb/yr. This is an **increase approaching 10 times**. |
| The **acceleration of the rate of increase** over the 15 years was calculated to be **15% per year, acceleration**. | The **acceleration of the rate of increase** over the 15 years was calculated to be **15% per year, acceleration, demonstrating non-linear or exponential growth.**.<br><br>**Methane's rate of increase is dramatically greater than $CO_2$, which is almost linear: $.28 x^2$ vs $.017 x^2$ for $CO_2$. Methane's variability at Barrow, is dramatically greater than at Mauna Loa, with 2024 hourly variations 5 times greater, and annual range 3 times greater.** |

**Diel Variability -**

$CO_2$ - The variability of $CO_2$ was not analyzed in detail, due to the lower variability (Fig's S4 and S5) and the modest warming potential when compared to methane.

**$CH_4$** – Methane variability was analyzed. Hourly variability is relevant for a highly-dynamic emissions-generating environment, which we contend is the present emissions situation. If hourly variations are substantially greater than weekly, monthly or annual variations, the use of multi-year warming potential such as GWP, or GWP*, over long-term time horizons may greatly misrepresent the atmospheric changes, and concomitant temperatures, and possibly extreme temperature excursions.

Typical seven-day emissions were plotted for MLO and BRW, in January and June, revealing high hourly variations, Figures S8 for MLO, and S9 and S10 for BRW. Two aspects are important:
-That modelling temperatures using long time horizon data will not be accurate, and
- That the dynamics of methane emissions generation are temperature dependent (Conrad, 2023), a positive, and possibly strong, "bursts" of methane, with a short-term, feedback loop, with possible micro-climate methane "domes" further accelerating additional methane emissions, on an hourly basis.

**Impact on Warming Determination.** June variability, 80 ppb, is substantially greater than January of 25 ppb: over three times. These oscillations of methane concentration, prevent attempting to accurately determine the warming effect by modelling. Using a metric such as GWP or GWP* is also particularly difficult, as they use longer time horizons of decades.

**Variability of $CH_4$ Emissions -** hourly, daily, seasonally. The variability of methane emissions is particularly relevant for a non-linear, exponential, and temperature-related system, particularly when biological emissions of methane are considered. This applies to waste decomposition, melting permafrost, and shallow, wet soils and ponds. Biological systems generally respond exponentially to rising temperatures, doubling every 5 to 10 $^{o}$C. Methane generation in anaerobic sediments at 5 $^{o}$C will be minimal but at 40 $^{o}$C methane production would be more than 100 times higher (Conrad, 2023). In July 2023 in the sub-Arctic area of Normal Wells, NT, Canada (Lat 65$^{o}$,16'N), the temperature reached 37.9 $^{o}$C, 15 $^{o}$ C above the normal July mean daily maximum. Such temperature extremes are coinciding with substantial increases in methane generation from the Arctic and sub-Arctic.

**Methane "Bursts" of Emissions** - Based on our data analyses, we suggest that some methane emissions, particularly for rapid emissions increases, may be responses to temperature increases, or other factors, possibly in combination, such that in the case of Barrow Alaska, the emissions occur to a large extent, as a series of short "bursts", not a homogenous or constant flow of methane, slowly varying or increasing with time as assumed in GWP analyses. Such dynamic and variable emissions cannot be modelled using long-time horizons.

With increasing global temperatures, increasing methane emissions can be expected, from positive temperature feedback. In August 2020, the principal researcher experienced warm, water-saturated soils, subsequent to spring flooding , latitude 51.5$^{o}$ N, "bubbling" with gas emissions. Recent temperatures had been unseasonably high, from 26.8 $^{O}$ C to 33.9 $^{O}$ C, from mid-June to mid-August. The "bubbling" had not previously been observed. Due to hazards of possible fire spreading, no attempt was made to ignite the gases observed in order to determine if methane was present. However, higher temperatures inevitably lead to greater methane emissions in damp soils.

Mauna Loa: To illustrate methane's variability, with location (latitude), and time of year, a "week" or seven day plot of methane emissions at Mauna Loa was made, January 1 through 7, 2024, Figure S3 below. The substantial hourly variation is evident on several days of 30 ppb, which is 1.5% of the concentration, but is 1/3 of the total annual variation, in a single hour. Daily variation was 33 ppb, while seven day was 34 ppb, essentially the same. This variation was 33% of annual variation.

It is notable that peaks increase from noon onward, and it is postulated that this corresponds with temperature increases, as peaks occur from 2pm to 6pm.

We have not attempted to explain causality, particularly with methane, however, using observations of variability has led us to contend that methane emissions appear to be
temperature related.  The principal implication is that anthropogenic emissions and resulting global warming create a positive feedback loop, resulting in further warming.  Thus reducing methane emissions is an essential action, requiring effective policies for implementation, including more rapid response to wildfires, reducing forest harvesting waste and decomposition, and other methane sources.

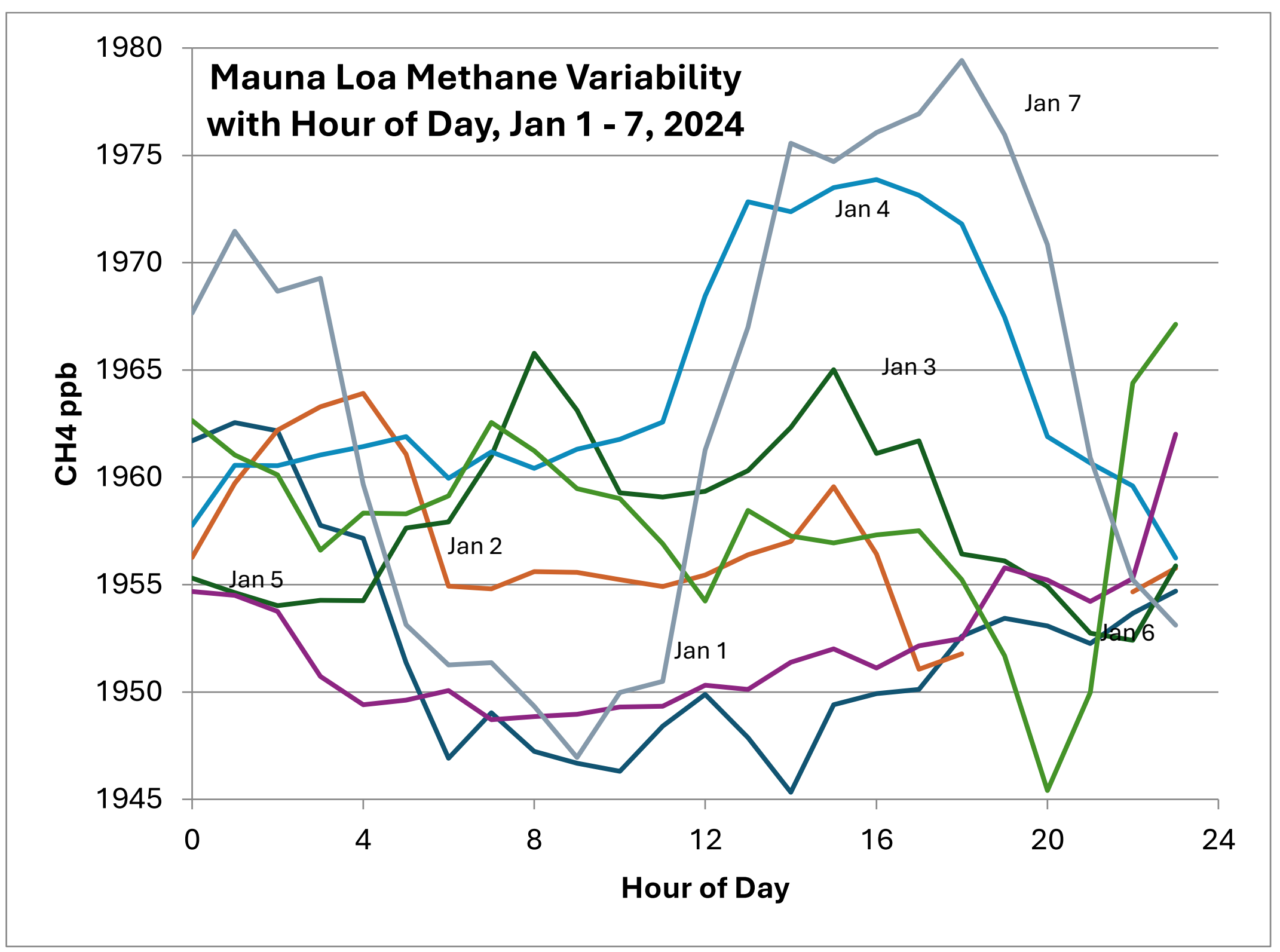

Figure S8.  Hourly methane emissions at Mauna Loa, HI, Jan 1 through 7th, 2024.     Bryenton, 2025

**$CH_4$ Variation Summary – ppb and %, MLO.**

<u>Hourly:</u>   Jan 6 – 14ppb/hr, 13.7% of annual variation and .7% of concentration; Jan 7 -10 ppb/hr, 9.8% of annual.

<u>Daily</u>: Jan 7 was 33 ppb/day, 27.5% of annual and 1.7% of concentration; Jan 4 was 18 ppb/day, 15% of annual variation.

<u>Seven day:</u> The variation range was 34 ppb, or 33% of annual variation.

<u>Peaks</u> – increase from mid morning to peak late day. Lows mid-morning to late am.  Plausible phase shift with sun and daily temperature increase.

**$CH_4$ Variation Summary – ppb and %, BRW.**

Barrow: A seven-day plot of Barrow was made for January 1 through 7, 2024, Figure S9; and for June 14 through 21, 2024, Figure S10, to illustrate latitude and seasonal differences between MLO and BRW. January BRW hourly variation was 4 ppb/hr, while June was 20 ppb, more than five times greater. Daily January variation was 12 ppb/day, while June was 51 ppb/day, more than four times greater. Seven day variation was 81 ppb variation which was 21% of annual variation. This variation of 81 ppb was 2 ½ times that of Mauna Loa.

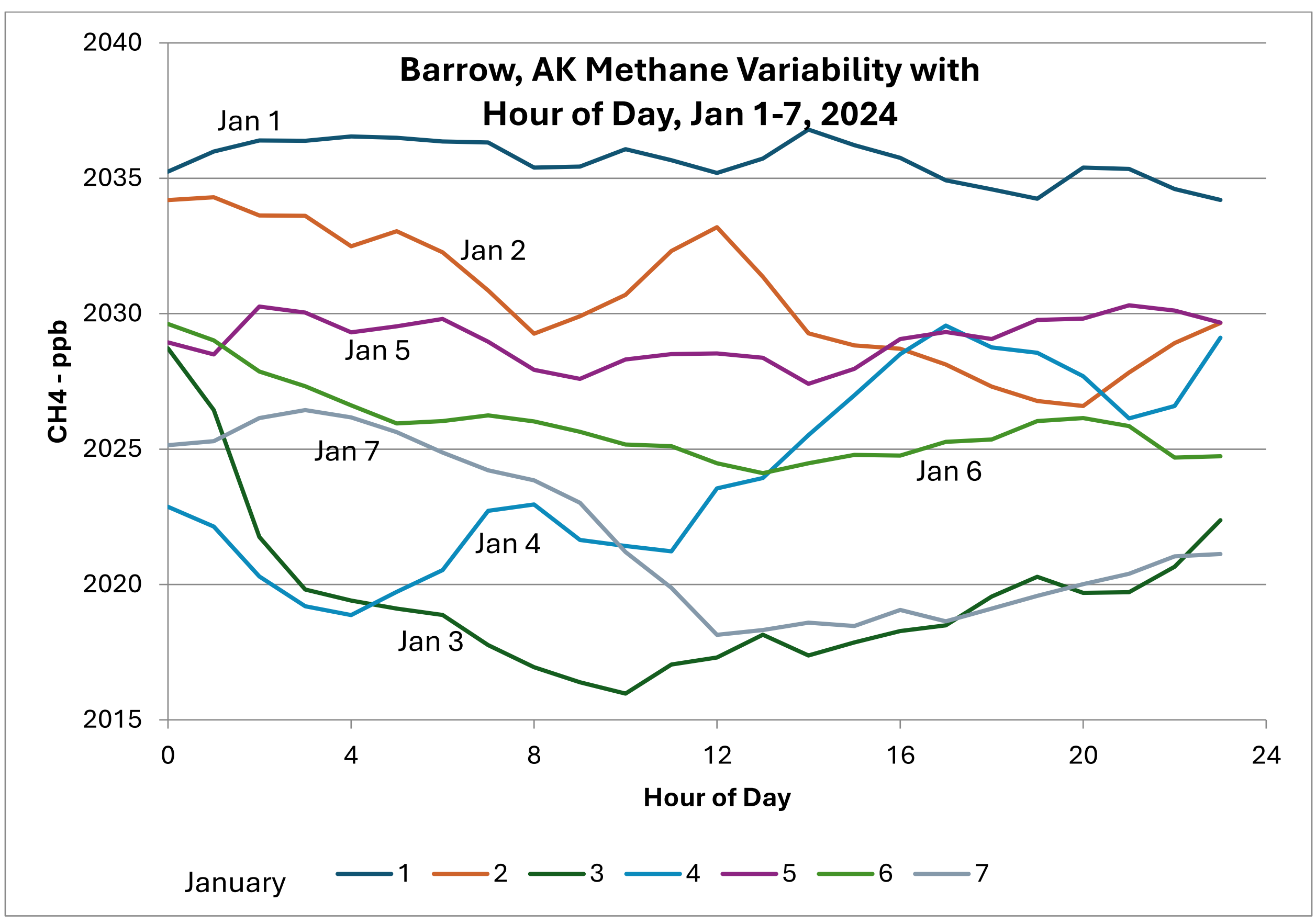

Figure S9. Barrow Alaska, Hourly Methane Emissions, Jan 1 – 7, 2024 Bryenton, 2025

**.January 1-7, 2024**

Hourly: Jan 2, 3 and 7 have more than 4 ppb/hr, 1.3% of annual variation .2% of concentration.

Daily: Jan 3 had 12ppb/day, 4% of annual variation and .6 % of concentration.

Seven day: seven-day (weekly) variation was 20 ppb or 6.8% of annual variation.

Due to the extreme annual variation of 378 ppb, or 18% of concentration, the hourly, daily and seven-day variations as percentage of annual variation and concentration appear much lower than Mauna Loa.

Further, the winter season in Barrow, being much lower temperature than Mauna Loa, would be expected to have substantially lower methane emissions, due to reduced biological activity.

Peaks – increase from mid-day to late day, early am. Plausible phase shift with sun and temperature. Lows early to late am.

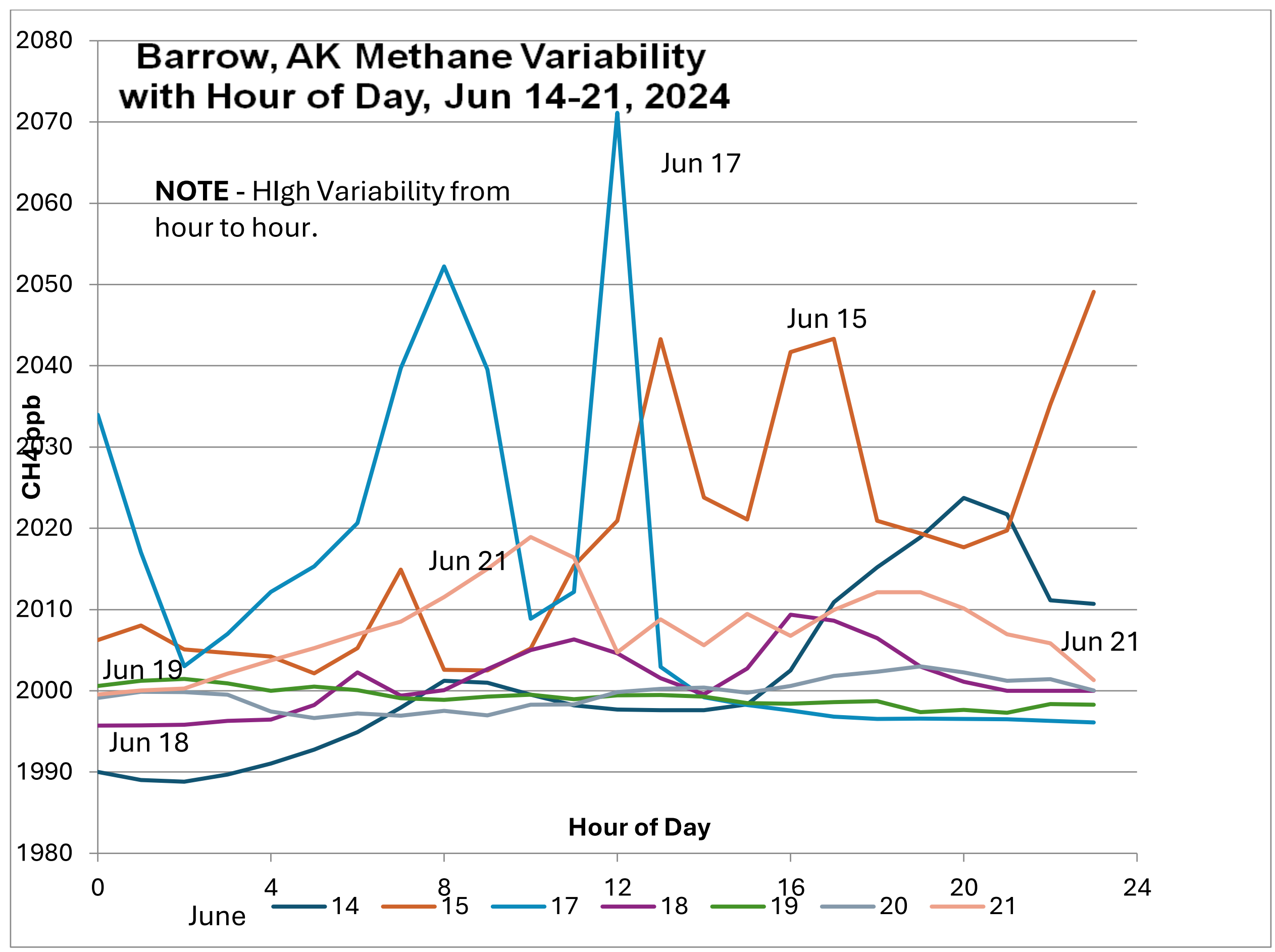

Figure S10. Barrow Alaska, Hourly Methane Emissions, June 14 – 21, 2024. June 16 deleted due to extreme, anomalous value of 2250 ppb and possible equipment malfunction Bryenton, 2025

**$CH_4$ Variation – ppb and %, BRW, June 14 – 21, 2024.**

Hourly - June 15 and 17 have more than 20 ppb/hr, or 6.8 % of annual variation.
Daily: June 15 has 51 ppb/day, 17.3 % of annual variation, and 2.5 % of concentration.
Seven-day: seven-day (weekly) variation was 81 ppb, or 27.6% of annual variation.

Peaks – increase from mid morning to peak late day. Lows early to late am. Plausible phase shift with sun and temperature.

**Observed Emissions** - With increasing global temperatures, future increases of methane emissions can be expected, from positive temperature feedback.